\documentclass[aps, prd, twocolumn, showpacs, superscriptaddress, groupedaddress, floatfix]{revtex4-2}
\usepackage{caption}
\usepackage{graphicx}
\usepackage{yfonts}
\usepackage{float}
\usepackage{amsmath}
\usepackage{amssymb}
\usepackage{amsfonts}
\usepackage{amsthm}
\usepackage{mathtools}
\usepackage{derivative}
\usepackage{dcolumn}
\usepackage[dvipsnames]{xcolor}
\usepackage{listings}
\usepackage{subfigure, rotating, bm, array}
\usepackage[utf8]{inputenc}
\usepackage[english]{babel}
\usepackage[normalem]{ulem}
\usepackage[pagebackref=false, colorlinks=true]{hyperref}
\hypersetup{
linkcolor=blue,     
citecolor=blue,     
urlcolor=blue}      

\usepackage{float}

\begin{document}
\title{End equilibrium state of a spherical gravitational collapse in the presence of matter and scalar field}
\author{Debanjan Debnath}
\email{debjanjan@gmail.com}
\affiliation{Department of Physics, Indian Institute of Technology, Kanpur, Uttar Pradesh 208016, India}
\author{Dipanjan Dey}
\email{deydipanjan7@gmail.com}
\affiliation{Beijing Institute of Mathematical Sciences and Applications, Beijing 101408, China}
\author{Kaushik Bhattacharya}
\email{kaushikb@iik.ac.in }
\affiliation{Department of Physics, Indian Institute of Technology, Kanpur , Uttar Pradesh 208016, India}
\date{\today}
\begin{abstract}
We explore the possibilities of modeling a spherically symmetric static spacetime that can emerge as the end state of gravitational collapse, by considering it to be seeded by a composite fluid made of matter and a scalar field. In this scenario, the matter represents dark matter, while the scalar field represents dark energy. On certain scales, dark energy is believed to significantly influence the structure formation of dark matter. Various models describe the possible impacts of dark energy on structure formation under different scenarios.
By investigating an inhomogeneous scalar field representing dark energy, coupled with dark matter, we demonstrate that this two-component fluid can seed spacetimes forming the final equilibrium state. We derive solutions for the scalar field and potential for Joshi-Malafarina-Narayan (JMN) spacetimes.
\end{abstract}
\maketitle

\section{Introduction}
\label{sec0}
Understanding the dynamic evolution and the intrinsic nature of cosmological structures at the galactic or galaxy cluster scale have been a major focus of research over the past several decades. This quest for knowledge may unravel some deeper insight related to structure formation in the near future.  Numerous empirical proposals have been suggested regarding structure formation, primarily based on observational data such as galactic rotation curves and gravitational lensing. These proposals have been tested with varying degrees of success. It is well-accepted from various indirect astrophysical and cosmological observations that the basic constituent of the cosmological structures is dark matter.
Quantum fluctuations during the inflationary era are believed to be responsible for the large-scale structure of the universe by causing matter density perturbations \cite{LiddleLyth93, white, Sengor18}. These perturbations initially grow within the framework of linear theory until they become nonlinear and form over-dense regions, primarily composed of dark matter. Since dark matter decouples from the primordial soup earlier than baryonic matter, the primordial density perturbations grow first in dark matter fluid resulting in over-dense regions of dark matter. In the late universe these regions are modeled by the `top-hat collapse model' \cite{GunnGott72}, which assumes spherical symmetry and uses a closed Friedmann-Lemaitre-Robertson-Walker (FLRW) metric.
Initially expanding with the background FLRW spacetime, these over-dense regions eventually collapse under gravity. General relativity predicts that such a collapse results in black holes \cite{OppenheimerSnyder39, Datt, Joshi07}. Therefore, to describe the formation of galactic halos, Newtonian virialization techniques are employed. 

Beyond a certain cosmological scale, the impact of dark energy on structure formation becomes significant. Several studies have explored how the dynamics of over-dense regions differ from those predicted by the top-hat collapse model. Dark energy may play a crucial role in the total Newtonian gravitational potential of over-dense regions \cite{Maor:2005hq, Saha1}. Models of structure formation in the presence of dark energy can be classified based on its contribution to the total gravitational potential:
\begin{itemize}
    \item The isolated sub-universe model neglects the influence of dark energy, akin to the top-hat collapse model.
    \item In the homogeneous dark energy model, the density of dark energy remains uniform across the boundary of over-dense regions \cite{Lahav:1991wc, Steinh, Shapiro, Horellou:2005qc}.
    \item The clustered dark energy model allows dark energy to cluster inside over-dense regions \cite{Basilakos:2003bi, Maor:2005hq, Basilakos:2006us, Basilakos:2009mz, Chang:2017vhs, Dey2, Saha2}.
    \item The clustered and virialized dark energy model considers the virialization of clustered dark energy \cite{Maor:2005hq}.
\end{itemize}
In \cite{Saha1} and \cite{Saha2}, a general relativistic approach is taken up to virialization to model homogeneous and clustered dark energy, respectively. These models consider quintessence and phantom-like scalar fields along with dust-like dark matter to represent the resultant fluid in over-dense regions. 

In \cite{Dey2, Dey3}, the authors use a fully general relativistic approach to model the dynamics of over-dense regions, where an end equilibrium state is defined covariantly. The scale factors ($a(r,t)$) of the over-dense patches should have the following constraint so that the over-dense regions can reach the equilibrium configuration in the comoving time $t_e$: 
\begin{eqnarray}
   a(r,t) &>& a_e(r)~ \forall t\in [0,t_e) \implies \dot{a} < 0 ~ \forall t\in [0,t_e)\,\, ,\nonumber\\
 \text{and}~  a(r,t) &=& a_e(r)~ \forall t\geq t_e \implies \dot{a} = \ddot{a} = \dddot{a} =...\nonumber \\&=& a^{(n)} = 0 ~ \forall t\geq t_e \,\, ,
 \label{equil}
\end{eqnarray}
where $a_e(r)$ is the value of the scale factor at the equilibrium time $t_e$. If the collapsing system reaches equilibrium at a finite comoving time ($t_e$), the first derivative of the scale factor $a(r, t)$ would be discontinuous ($\mathcal{C}^0$) at $t_e$. However, since we generally assume $g_{\mu\nu}(t,r)$ is at least $\mathcal{C}^2$, this is not possible. Therefore, equilibrium can only be achieved at asymptotic comoving time (i.e., $t_e\to \infty$). In \cite{Dey3}, it is demonstrated that a homogeneous, spherically symmetric over-dense patch consisting of a scalar field with a non-zero potential can reach the aforementioned equilibrium configuration if the potential has certain specific forms. However, it should be noted that the potential derived in that paper as a function of the scalar field is only valid near the equilibrium state.
In a cosmologically relevant context \cite{Koushiki1}, using a similar mathematical technique, the authors show that a spherically symmetric over-dense region, consisting of minimally coupled dust-like dark matter and a scalar field acting as dark energy, can reach equilibrium due to specific forms of the scalar field's potential. This potential, with its specific forms, generates non-zero negative pressure inside the collapsing over-dense regions, leading to the final equilibrium state. 

A natural extension of the collapse model involving a homogeneous scalar field minimally coupled with homogeneous dust-like matter introduces inhomogeneity into the collapsing matter. In \cite{Joshi1}, the authors demonstrate that certain spherically symmetric, Petrov type-D inhomogeneous solutions of the Einstein equations, known as Joshi-Malafarina-Narayan (JMN) spacetimes, can represent the end equilibrium state of a collapsing inhomogeneous spherically symmetric compact object. This raises the question of whether these types of spacetimes, shown to form as an end equilibrium state of gravitational collapse, can be modeled by scalar fields and dust-like matter. In this paper, we explore this problem. 

It can be shown that if a static spherically symmetric spacetime is seeded solely by a time-independent inhomogeneous scalar field $\varphi(r)$, then the spacetime should satisfy the following forms of the components of the energy-momentum tensor:
\begin{eqnarray}
  T^0_{0} &=&T^2_{2}=T^3_{3}=-\frac{e^{-2\beta(r)}}{2} {\varphi^{\prime}}^2 - V(\varphi),\nonumber \\
T^1_{1} &=& \frac{e^{-2\beta(r)}}{2}{\varphi^{\prime}}^2 - V(\varphi),
\label{basictmn}
\end{eqnarray}
where prime over any quantity denotes the partial derivative with respect to $r$ and the $(1,1)$ metric tensor component of the spherically symmetric static spacetime is $e^{2\beta(r)}$, where the line element is:
\begin{eqnarray}
    ds^2 =- e^{2\alpha(r)}dt^2+e^{2\beta(r)} dr^2+r^2d\Omega^2,
    \label{metric1}
\end{eqnarray}
where $\alpha(r)\, , \beta(r)$ are real-valued functions of radial coordinate $r$, and $d\Omega^2 = (d\vartheta^2 + \sin^2\vartheta d\phi^2)$.
Thus, a spacetime corresponding to an equilibrium configuration resulting from gravitational collapse must exhibit a restrictive type of stress-energy tensor if it is seeded only by a scalar field. However, if we consider a composite fluid consisting of both matter and a scalar field, this constraint on the stress-energy tensor is relaxed. Additionally, this scenario is more relevant in a cosmological context where dark energy significantly influences the structure formation of dark matter beyond a certain length scale. 
In this paper, we examine an inhomogeneous scalar field $(\varphi(r))$, representing dark energy, which is coupled (either minimally or non-minimally) with dark matter\footnote{Recently there are many papers which deal with non-minimal coupling of canonical or non-canonical scalar fields with dark matter as can be seen in Refs.~\cite{Hussain:2023kwk, Hussain:2022dhp, Hussain:2022osn, Chatterjee:2021ijw}.}. We demonstrate that this resultant two-component fluid can seed spacetimes that potentially form as the final equilibrium state of gravitational collapse. As an illustration, we derive the solutions for $\varphi(r)$ and $V(\varphi)$ for the first and second kinds of JMN solutions (JMN-1 and JMN-2 solutions, respectively). The results for the JMN-1 spacetimes exhibit remarkable similarities to those obtained in \cite{Dey3} concerning the form of the potential $V$ as a function of $\varphi$. These similarities suggest that, in certain limited cases, the form of the scalar field potential responsible for the equilibrium state remains unchanged, even though the two-fluid system is inhomogeneous. In this scenario, the potential becomes a function of the radial coordinate, reflecting the scalar field's inhomogeneity. However, it should be noted that this paper does not explore the evolution of the inhomogeneous scalar field up to the equilibrium state. Instead, we focus on the solutions of a scalar field coupled with matter for the spherically symmetric static spacetimes that can form as the end state of gravitational collapse.

The work in this paper is organized in the following way. In section \ref{two}, we derive the expressions of a scalar field and its potential for a spherically symmetric static spacetime, where we consider both the minimal and non-minimal coupling between the matter and scalar field. In Section \ref{three}, we use the results in section \ref{two} to derive the scalar field solutions of JMN-1, JMN-2 spacetimes. Section \ref{four} gives a summary of the work presented in this paper. Throughout the paper, we use a system of units in which the velocity of light and the universal gravitational constant (multiplied by $8\pi$), are both set equal to unity.

\section{Solutions of scalar field coupled with isotropic matter for spherically symmetric, static spacetimes}
\label{two}
In this section, we explore the properties of a two-fluid system composed of a scalar field and isotropic matter, which seeds the general spherically symmetric static spacetime described by Eq.~(\ref{metric1}). Here, we consider matter with isotropic pressure and derive the corresponding solutions of $\varphi$. We start with the action \cite{Saha2}:
\begin{eqnarray}
\mathcal{S}=\int d^4x~\left(\mathcal{L}_{\text{grav}}+\mathcal{L}_{\text{m}}+\mathcal{L}_{\varphi}+\mathcal{L}_{\text{int}}\right)\,\, ,
\end{eqnarray}
where the gravitational sector is given by the standard Einstein–Hilbert Lagrangian:
\begin{eqnarray}
\mathcal{L}_{\text{grav}}=\frac{\sqrt[•]{-g}R}{2}\,\, ,
\end{eqnarray} 
where $g$ is the determinant of the metric tensor $g_{\mu\nu}$ and $R$ is the Ricci scalar.
The Lagrangian for the relativistic fluid is given by
\begin{eqnarray}\label{1}
\mathcal{L}_{\text{m}}=-\sqrt[]{-g}\rho_{m}(n,s)+J^{\mu}\left(\varkappa_{,\mu}+s\theta_{,\mu}+\beta_{A}\alpha^{A}_{,\mu}\right)\,\, ,
\end{eqnarray}
 where $\rho_{m}$ is the energy density of the matter. We assume $\rho_{m}(n,s)$ to be prescribed as a function of $n$, the particle number density, and $s$, the entropy density per particle. $\varkappa$, $\theta$, and $\beta_{A}$ are all Lagrange multipliers with A taking the values 1, 2, 3, and $\alpha^{A}$ are the Lagrangian coordinates of the fluid.  These Lagrange multipliers are introduced because of the various thermodynamic constraints followed by the fluid. In this paper we will not require the thermodynamic aspects and consequently we will not discuss about these multipliers in detail. 
 The vector density or the current density of particle number $J^{\mu}$ is related to $n$ as
$$J^{\mu}=\sqrt[]{-g}nU^{\mu},~ |J|=\sqrt[]{-g_{\mu\nu} J^\mu J^\nu},~ n=\frac{|J|}{\sqrt[]{-g}}\, ,$$
where $U^{\mu}$ is the timelike 4-velocity of matter satisfying $U_{\mu}U^{\mu}=-1$.
The scalar field Lagrangian is given by
\begin{eqnarray}\label{2}
\mathcal{L}_{\varphi}=-\sqrt[]{-g}~\left[\frac{1}{2}\epsilon\partial_{\mu}\varphi \partial^{\mu}\varphi+V(\varphi)\right]\, ,
\end{eqnarray}
where $\epsilon=1,-1$ are for quintessence and phantom-like scalar field, respectively and $V(\varphi)$ is the potential of the scalar field $\varphi$. Lastly, the Lagrangian for the interacting sector is
\begin{eqnarray}\label{3}
\mathcal{L}_{\text{int}}=-\sqrt[]{-g}f(n,s,\varphi)\, ,
\end{eqnarray}
where $f(n,s,\varphi)$ is an arbitrary function of $n,~s$ and $\varphi$.
The energy-momentum tensor, for any constituent sector, can be written as:
\begin{eqnarray}\label{4}
T_{\mu\nu}=\frac{-2}{\sqrt[]{-g}}\frac{\delta \mathcal{L}}{\delta g^{\mu\nu}}.
\end{eqnarray}
From the action one can find out the Einstein equation for the system in terms of the total energy-momentum tensor $T^\mu_\nu$. This energy-momentum tensor contains the contributions from the various  components. The energy-momentum tensors for each component can be obtained from the above equation. The energy density and pressure of the various components then can be obtained as
\begin{eqnarray}
T^0_0 = -\rho \,,\,\,\,\,\,T^i_j = p^{(i)} \delta^i_j\,,
\label{rpdef}
\end{eqnarray}
where in the above equation there is no sum over the index $i$. We use the above definition to calculate the pressure and energy density of an isotropic as well as an anisotropic fluid. For an isotropic fluid we have $p^{(i)} = p$, for every $i$.

Using the expression for the general spherically symmetric static spacetime (Eq.~(\ref{metric1})), and the corresponding Einstein equations, we can write:
\begin{eqnarray}
\label{E1}
   T^{0}_{0} &=& e^{ - 2\beta} \left( \frac{1}{r^2} - \frac{2\beta^{\prime}}{r} \right) - \frac{1}{r^2}, \\ 
   \label{E2}
   T^{1}_{1} &=&  e^{ - 2\beta} \left(\frac{2\alpha^{\prime}}{r} + \frac{1}{r^2} \right) - \frac{1}{r^2}, \\ 
   \label{E3}
   T^{2}_{2} &=&  T^{3}_{3} =  e^{- 2\beta} \left( \alpha^{\prime \prime} + {\alpha^{\prime}}^2 - \alpha^{\prime} \beta^{\prime} + \frac{\alpha^{\prime} - \beta^{\prime}}{r} \right).
\end{eqnarray}
In our model, the total energy-momentum tensor \(T_{\mu\nu}\) is composed of three components:
\begin{equation}
    T_{\mu \nu} = T^{(\text{m})}_{\mu \nu} + T^{(\text{int})}_{\mu \nu} + T^{(\varphi)}_{\mu \nu}. \label{ttot}
\end{equation}
Here, \(T^{(\text{m})}_{\mu \nu}\), \(T^{(\text{int})}_{\mu \nu}\), and \(T^{(\varphi)}_{\mu \nu}\) are the energy-momentum tensors corresponding to the matter part, the interaction part (between the matter fluid and the scalar field), and the scalar field, respectively. These tensors are given by:
\begin{eqnarray}
    T^{(\text{m})}_{\mu \nu} &=& p_{m} g_{\mu\nu} + (\rho_{m} + p_{m}) U_{\mu} U_{\nu}, \\
    T^{(\text{int})}_{\mu \nu} &=& p_{\text{int}} g_{\mu\nu} + (\rho_{\text{int}} + p_{\text{int}}) U_{\mu} U_{\nu}, \\
    T^{(\varphi)}_{\mu \nu} &=& \epsilon \partial_{\mu} \varphi \partial_{\nu} \varphi - g_{\mu \nu} \left\{ \frac{1}{2} \epsilon \partial_{\lambda} \varphi \partial^{\lambda} \varphi + V(\varphi) \right\},
\end{eqnarray}
where \(\epsilon = 1\) for quintessence and \(\epsilon = -1\) for a phantom-like scalar field. As the matter part is present in a static spacetime we assume $U^\mu = \left(e^{-\alpha}, 0, 0, 0\right)$.
Using the above equations we write Eq.~\eqref{ttot} as
\begin{widetext}
\begin{eqnarray}
    T_{\mu \nu} &= (p_{m} + p_{\text{int}}) g_{\mu\nu} + (\rho_{m} + \rho_{\text{int}} + p_{m} + p_{\text{int}}) U_{\mu} U_{\nu} + \epsilon \partial_{\mu}\varphi \partial_{\nu} \varphi - g_{\mu \nu}\left\{ \frac{1}{2} \epsilon \partial_{\lambda}\varphi \partial^{\lambda} \varphi + V(\varphi) \right\} \label{ttot2} \\ \nonumber 
    &= p g_{\mu\nu} + (\rho + p) U_{\mu} U_{\nu} + \epsilon \partial_{\mu}\varphi \partial_{\nu} \varphi - g_{\mu \nu}\left\{ \frac{1}{2} \epsilon \partial_{\lambda}\varphi \partial^{\lambda} \varphi + V(\varphi) \right\}\, ,
\end{eqnarray}
\end{widetext}
where we have denoted 
\begin{eqnarray}
p = p_{m} + p_{\text{int}}\,,\,\,\,\,{\rm and}\,\,\,\,\rho = \rho_{m} + \rho_{\text{int}}\,. 
\end{eqnarray}
Since we consider spherically symmetric static spacetimes, the components of the energy-momentum tensor are independent of temporal (i.e., $t$) and angular coordinates (i.e., $\theta$ and $\phi$). Therefore, the scalar field $\varphi = \varphi(r)$, and the components of the energy-momentum tensor become:
\begin{eqnarray}
T^0_{0} &=& - \rho  - \frac{1}{2} \epsilon e^{- 2\beta} {\varphi^{\prime}}^2 - V(\varphi),\label{E4} \\
T^1_{1} &=& p + \frac{1}{2} \epsilon e^{- 2\beta} {\varphi^{\prime}}^2 - V(\varphi),\label{E5} \\
T^2_{2} &=& T^3_3 = p - \frac{1}{2} \epsilon e^{- 2\beta} {\varphi^{\prime}}^2 - V(\varphi)\label{E6},
\end{eqnarray}
where we employ the isotropic nature (i.e., $T^{1(\text{m})}_1= T^{2(\text{m})}_2 = T^{3(\text{m})}_3$) of the energy-momentum tensor of the matter part.
In addition, the scalar field $\varphi$ satisfies the following Klein-Gordon equation:
\begin{eqnarray}
  \epsilon  \square \varphi - \frac{\partial V(\varphi)}{\partial \varphi} - \frac{\partial f(\rho_m, \varphi)}{\partial \varphi}=0\, ,
\end{eqnarray}
which takes the following form for the aforementioned static and spherically symmetric spacetime (Eq.~(\ref{metric1})):
\begin{eqnarray}
   \epsilon e^{- 2\beta} \left\{ \varphi^{\prime \prime} + (\alpha^{\prime} - \beta^{\prime}) \varphi^{\prime} + \frac{2}{r} \varphi^{\prime} \right\} - \frac{\partial V(\varphi)}{\partial \varphi} - \frac{\partial f(\rho_m, \varphi)}{\partial \varphi} = 0.\nonumber\\ \label{phieom}
\end{eqnarray}
Here, $f(\rho_m, \varphi)$ is the interaction term between matter and scalar field \cite{Saha2}. 
Now, using the Eqs.~(\ref{E1}, \ref{E2}, \ref{E3}) and the expressions of $T^0_0, T^1_1, T^2_2$ written in  Eqs.~(\ref{E4}, \ref{E5}, \ref{E6}), we can write:
\begin{eqnarray}
  T^0_0 + T^1_1 &=&  p - \rho - 2 V(\varphi) =  2e^{- 2\beta} \left( \frac{1}{r^2} + \frac{\alpha^{\prime} - \beta^{\prime}}{r} \right) - \frac{2}{r^2},\nonumber\\
  \label{ab}\\
  T^2_2 - T^0_0 &=& p + \rho\nonumber\\ &=&  e^{- 2\beta} \left( \alpha^{\prime \prime} + {\alpha^{\prime}}^2 - \alpha^{\prime} \beta^{\prime} + \frac{\alpha^{\prime} + \beta^{\prime}}{r} - \frac{1}{r^2}\right) + \frac{1}{r^2},\nonumber\\
  \label{cd}\\
   T^1_1 - T^2_2 &=&  \epsilon e^{- 2\beta(r)} {\varphi^{\prime}}^2 \nonumber\\&=& -e^{- 2\beta} \left( \alpha^{\prime \prime} + {\alpha^{\prime}}^2 - \alpha^{\prime} \beta^{\prime} - \frac{\alpha^{\prime} + \beta^{\prime}}{r} - \frac{1}{r^2}\right) - \frac{1}{r^2}. \nonumber\\
   \label{2e3}
\end{eqnarray}
Taking the positive root of $\varphi^{\prime}$ from Eq.~\eqref{2e3}, we obtain the following form of $\varphi$ after an integration:
\begin{eqnarray}
    \varphi(r) = C_1 + \int dr \hspace{1mm}e^{\beta(r)} \sqrt{\epsilon\left[T^1_1 - T^2_2 \right]}, \label{phi}
\end{eqnarray}
where $C_1$ is the integration constant. The above expression of $\varphi$ indicates that $\varphi$ becomes constant when the composite fluid is isotropic in nature (i.e., $T^1_1 = T^2_2$). Now, using the Klein-Gordon equation of $\varphi$ (Eq.~(\ref{phieom})) and the expression of $\varphi^\prime$, we can write:   
\begin{widetext}
    \begin{align*}
    \epsilon \frac{\partial}{\partial \varphi} (V + f) &= e^{- 2\beta} \left\{ \varphi^{\prime \prime} + (\alpha^{\prime} - \beta^{\prime}) \varphi^{\prime} + \frac{2}{r} \varphi^{\prime} \right\} \\
    &= e^{- 2\beta} \left\{ \beta^{\prime} e^{\beta} \sqrt{\epsilon\left(T^1_1 - T^2_2\right)} + e^{\beta} \frac{\epsilon\left({T^{1}_1}^{\prime} - {T^{2}_2}^{\prime}\right)}{2\sqrt{\epsilon\left(T^1_1 - T^2_2\right)}} + \left(\alpha^{\prime} - \beta^{\prime}\right) e^{\beta} \sqrt{\epsilon\left(T^1_1 - T^2_2\right)}
     + \frac{2}{r} e^{\beta} \sqrt{\epsilon\left(T^1_1 - T^2_2\right)} \right\} \\
     &= e^{-\beta} \left\{ \beta^{\prime}  \sqrt{\epsilon\left(T^1_1 - T^2_2\right)} +\frac{\epsilon\left({T^{1}_1}^{\prime} - {T^{2}_2}^{\prime}\right)}{2\sqrt{\epsilon\left(T^1_1 - T^2_2\right)}} + (\alpha^{\prime} - \beta^{\prime}) \sqrt{\epsilon\left(T^1_1 - T^2_2\right)}
     + \frac{2}{r} \sqrt{\epsilon\left(T^1_1 - T^2_2\right)} \right\}.
\end{align*}
Multiplying the above equation by $\varphi^{\prime}$ we get
\begin{align*}
    \epsilon \frac{d V}{d r} + \epsilon \frac{\partial f}{\partial \varphi} \varphi^{\prime}  &= \left\{ \beta^{\prime}  \epsilon\left(T^1_1 - T^2_2\right) +\frac{1}{2}\epsilon\left({T^{1}_1}^{\prime} - {T^{2}_2}^{\prime}\right) + (\alpha^{\prime} - \beta^{\prime}) \epsilon\left(T^1_1 - T^2_2\right)
     + \frac{2}{r} \epsilon\left(T^1_1 - T^2_2\right) \right\}, 
\end{align*}
which  after a rearrangement gives
\begin{eqnarray}
     \frac{d V}{d r} + \frac{\partial f}{\partial \varphi} \varphi^{\prime}  &= \left\{ \left(\alpha^{\prime} + \frac{2}{r} \right) \left(T^1_1 - T^2_2\right) + \frac{1}{2} \left({T^{1}_1}^{\prime} - {T^{2}_2}^{\prime} \right) \right\}. 
\label{vpf}     
\end{eqnarray}
\end{widetext}
The function $f$, which specifies the field-matter interaction, can be of the following form:
\begin{eqnarray}
f(\rho_m,  \varphi) = \gamma \zeta(\rho_m) \Phi(\varphi)\,,
\label{mulf}    
\end{eqnarray}
which assumes that the interaction Lagrangian is simply the product of some function of the matter energy density, $\zeta(\rho_m)$, and a function of the scalar field, $\Phi(\varphi)$. This is a simple form of interaction. One may employ a different kind of interaction term where the interaction is additive, i.e., where a function of the scalar field is added to another function of the fluid energy density. We will work with such an interaction later. These in general are the simplest kind of non-minimal interactions. One can use more complicated interactions if needed.

We can now integrate the expression in Eq.~(\ref{vpf}), with proper boundary condition, to get an expression for $V + f$
\begin{eqnarray}
    V + f = C_2 + \int dr \left(\alpha^{\prime} + \frac{2}{r} \right) \left(T^1_1 - T^2_2\right) + \frac{1}{2} \left(T^1_1 - T^2_2 \right), \label{v+f} \label{v+fsolngen} \nonumber\\ 
\end{eqnarray}
where $C_2$ is the integration constant. Our formal introduction of the theory ends here. To proceed further we have to choose some specific spacetime solution of the Einstein equation. Moreover to work out the details we also require some specific scalar field potential. We introduce all these specific features in the next section. 

\section{Examples of scalar field solutions for spherically symmetric static spacetimes}
\label{three}
\subsection{JMN Spacetimes}

In \cite{Joshi1, Joshi2}, the authors present a class of static, spherically symmetric spacetimes that satisfy the equilibrium conditions at the end of the gravitational collapse of a spherically symmetric compact object. This collapse is modeled by the following metric:
\begin{eqnarray}
    ds^2 = -e^{2\nu(r,t)}dt^2 + \frac{{R^{\prime }}^2}{G(r,t)}dr^2 + R^2(r,t)d\Omega^2,
    \label{sptcollapse1}
\end{eqnarray}
where $\nu(r,t)$ and $G(r,t)$ are positive real-valued functions of the comoving coordinates $r$ and $t$, and $R(r,t)$ is the physical radius of a collapsing shell. The prime on $R(r,t)$ denotes the partial derivative with respect to $r$. The Einstein equations relate these functions and their derivatives to the energy density and pressures $\rho (r, t)$, $p^{(r)}(r, t)$, $p^{(\vartheta)}(r, t)$ of the system~\cite{Joshi3, Joshi8}. In general the number of total unknown functions are greater than the number of Einstein equations. This gives us the freedom to chose some of the unknown functions at our will.  The general scheme to formulate the collapse process is then, to evolve this system through the Einstein equations along with some specific choice of initial conditions (initial data) for the system. The initial data should be chosen such that certain regularity conditions are met: (i) the initial data should be sufficiently regular such that, these describe the viable physical conditions of the system at the beginning of the collapse. One such example is that, the total force on a particle is zero at the center of the matter cloud. This means that, initially  the gradient of the total pressures are all zero at the center, i.e., $\partial p^{(r)}/\partial r |_{r = 0} = 0 = \partial p^{(\perp)}/\partial r|_{r = 0}$. Here $p^{(\perp)}$ represents total tangential stress ($p^{(\vartheta)}$ or $p^{(\phi)}$), in case the system has anisotropic pressures.  Also from the physical standpoint, initially the difference between the total radial pressure and the total tangential pressure should be zero at the center, i.e., $p^{(r)}|_{r = 0} - p^{(\perp)}|_{r = 0} = 0$.  (ii) The unknown metric functions should be chosen, such that, the $\mathcal{C}^2$-differentiability is maintained, as required by the Einstein equations; (iii) the free functions should be chosen, such that, the required energy conditions are satisfied. In addition to these, the avoidance of shell crossing singularity (which arises when matter shells having different radial coordinates collide with each other) puts additional constraint: $R^{\prime}(r, t) > 0$, throughout the collapse process~\cite{Joshi4, Joshi7}. By proper choice of these free functions and initial conditions, we can have either a black hole or a naked singularity as the end state product of the gravitational collapse~\cite{Joshi9}. The matter field is chosen to be of type I (section 4.3 of ~\cite{LSS}), satisfying weak and dominant energy conditions and encompasses almost all relevant known type matter. The equilibrium configuration is achieved when $\nu(r, t)$, $G(r, t)$, and $R(r, t)$ depend solely on $r$ and satisfy specific equilibrium conditions derived from Eq.~\eqref{equil} in the limit of asymptotic comoving time. Under these considerations and using the constraints from Eq.~\eqref{equil}, the authors in \cite{Joshi1, Joshi2}, derive the metric components of a class of spherically symmetric static spacetimes known as Joshi-Malafarina-Narayan (JMN) spacetimes. It must be noted that the background universe is modeled as an expanding FLRW spacetime. At a smaller length scale, one can have an over-dense spherical patch that detaches from the universal expansion and collapses ultimately in an inhomogeneous medium as worked out in 
Ref.~\cite{Bhattacharya:2017chr}. This inhomogeneous medium seeds the above spacetime.  

There are two kinds of JMN spacetimes namely the JMN-1 and JMN-2 spacetimes and the following line elements define them~\cite{Joshi1, Joshi2, Joshi5, Joshi6}:
\begin{widetext}
\begin{eqnarray}
 ds^2_{JMN-1} &=& -(1- M_0) \left(\frac{r}{R_b}\right)^\frac{M_0}{(1- M_0)}dt^2 + \frac{dr^2}{(1 - M_0)} + r^2d\Omega^2 \,,
\label{JMN-1metric} \\
 ds^2_{JMN-2} &=& -\frac{1}{16\lambda^2(2-\lambda^2)}\left[(1+\lambda)^2\left(\frac{r}{R_b}\right)^{1-\lambda}-(1-\lambda)^2\left(\frac{r}{R_b}\right)^{1+\lambda}\right]^2dt^2 + \left(2-\lambda^2\right)dr^2 + r^2d\Omega^2 \,.
 \label{JMN2metric}
\end{eqnarray}
\end{widetext}
Both spacetimes match an external Schwarzschild geometry at $r = R_b$, and $M_0$ and $\lambda$, which are dimensionless constants, are always positive for JMN-1 and JMN-2, respectively. 
The JMN spacetimes can be matched with Schwarzschild spacetime at $r = R_b$. Therefore, the resulting spacetime configuration is internally JMN-1 or JMN-2 and externally Schwarzschild, where the external Schwarzschild spacetime can be written in the following form:

\begin{equation}
ds^2 = -\left(1-\frac{M_0R_b}{r}\right)dt^2 + \frac{dr^2}{\left(1-\frac{M_0R_b}{r}\right)} + r^2d\Omega^2 \,,
\label{schext}
\end{equation}
Here, the Schwarzschild radius $R_s = M_0 R_b$, and since \( 0 < M_0 \leq 4/5 \), \( R_b > R_s \). The total Schwarzschild mass \( M_{S} \) is \( \frac{M_0 R_b}{2} \).
In general relativity, the matching of two spacetimes at a specific spacelike or timelike hypersurface requires satisfying two junction conditions. First, the induced metrics of the internal and external spacetimes on the matching hypersurface must be identical. This condition is satisfied at $r = R_b$, giving $M_0 = \frac{1-\lambda^2}{2-\lambda^2}$ for JMN-2. Second, the extrinsic curvatures $K_{ab}$ at the hypersurface must match.
The zero radial pressure in JMN-1 ensures that the extrinsic curvatures of the JMN-1 and Schwarzschild spacetimes at $r = R_b$ match automatically. For JMN-2, the pressure must vanish at $r = R_b$ to match the extrinsic curvature. The pressure in JMN-2 is given by:
\begin{equation}
    p = \frac{1}{(2-\lambda^2)}\frac{1}{r^2}\left[\frac{(1-\lambda)^2 A - (1+\lambda)^2 B r^{2\lambda}}{A - B r^{2\lambda}}\right] \,,
\end{equation}

where $A = \frac{(1+\lambda)^2 R_b^{\lambda-1}}{4\lambda \sqrt{2-\lambda^2}}$ and $B = \frac{(1-\lambda)^2 R_b^{-\lambda-1}}{4\lambda \sqrt{2-\lambda^2}}$. It can be verified that $p$ becomes zero at $r = R_b$.

\subsubsection{Minimally coupled scalar field solutions for JMN-1 spacetimes}

In this case we assume the interaction between the scalar field and the matter sector vanishes and consequently $f=0$. As because non-minimal interaction vanishes we always expect $\rho_{\rm int}$, $p_{\rm int}$ are all individually zero. Comparing the line element of JMN-1 (Eq.~(\ref{JMN-1metric})) spacetime with the line element of static spherically symmetric spacetime given in Eq.~(\ref{metric1}), we get:
\begin{eqnarray}
    \alpha(r) &=& \frac{1}{2} \ln(1-M_0) + \frac{M_0}{2(1-M_0)} \ln\left(\frac{r}{R_b}\right), \nonumber\\ \beta &=& -\frac{1}{2} \ln(1-M_0).
\end{eqnarray}
The energy-momentum tensor components $T^0_0$, $T^1_1$ and $T^2_2$ for JMN-1 spacetime are given by
\begin{align}
    T^0_0 &= - \frac{M_0}{r^2}, \label{f0soln2} \\
    T^1_1 &= 0, \label{f1soln2} \\
    T^2_2 = T^3_3 &= \frac{M^2_0}{4r^2(1-M_0)}. \label{f2soln2}
\end{align}
From the above expressions and Eq.~(\ref{basictmn}) it is clear that a scalar field alone will not be able to seed the JMN-1 spacetime as for a scalar field $T^1_1$ is in general not equal to zero. The above equations say that a scalar field in conjunction with a matter part can seed the JMN-1 spacetime, granted that their $T^1_1$ components cancel each other. This can happen for the case where we have dark matter with a small isotropic positive pressure and a scalar field with negative radial pressure.  From Eq.~\eqref{phi} we can see that, as $T^1_1 - T^2_2$ is negative, we can have a real solution for $\varphi(r)$, only  for a phantom-like scalar field $\epsilon = -1$
\begin{eqnarray}
    \varphi(r) &=&  \int_{R_b}^{r} dr \hspace{0.5mm} e^{\beta(r)} \left[\epsilon\left(T^1_1-T^2_2\right)\right]^{1/2}\nonumber\\ &=& \frac{M_0}{2(1-M_0)} \ln \left(\frac{r}{R_b} \right)\,. \label{phisoln3}
\end{eqnarray}
In the above expression, we have integrated from $R_b$ to $r$ as $R_b$ naturally gives us a length scale. The expression of the solution of $\varphi(r)$ shows that it becomes zero on $R_b$. As we are working with a spacetime which has a central singularity, the scalar field blows up at the center.  

From Eqs.~\eqref{ab}, \eqref{cd}, \eqref{v+f}, we get
\begin{eqnarray}
    &p_m(r)  = V(\varphi) + \frac{M^2_0}{8(1-M_0)}\frac{1}{r^2}, \label{psoln3} \\
    &V(\varphi(r)) = C_2 + \frac{M^2_0(2-M_0)}{16(1-M_0)^2} \frac{1}{r^2}, \label{v+fsoln3} \\
    &\rho_m(r) = - C_2 + \frac{M_0(4-3M_0)(4-5M_0)}{16(1-M_0)^2}\frac{1}{r^2}. \label{rhosoln3}  
\end{eqnarray}
The above relations can be written in the following compact form in the minimally coupled case (i.e., the non-interacting case) as the following:
\begin{eqnarray}
V(\varphi) &=& \widetilde{V}_0 e^{-2 \varphi / \varphi_0} + C_2, \\
\rho_m(r) &=& \frac{\widetilde{\rho}_0}{r^2} - C_2, \\
p_m(r) &=&\frac{\widetilde{p}_0}{r^2} + C_2, 
\end{eqnarray}
where
\begin{eqnarray}
   \varphi_0 &=& \frac{M_0}{2(1-M_0)}, \\
   \widetilde{V}_0 &=& \frac{M^2_0(2-M_0)}{16 {R_b}^2 (1-M_0)^2}, 
   \label{vtild}\\
   \widetilde{\rho}_0 &=& \frac{M_0(4-3M_0)(4-5M_0)}{16(1-M_0)^2}, 
   \label{rtild}\\
   \widetilde{p}_0 &=& \frac{M^2_0(4-3M_0)}{16(1-M_0)^2}.
   \label{ptild}
\end{eqnarray}
The form of the scalar field potential obtained here shows a remarkable similarity to the potential of the homogeneous scalar field derived in \cite{Dey3, Koushiki1} to describe the equilibrium state. It is important to note that in those papers, the potential as a function of $\varphi$ is expressed at the equilibrium state limit. Another significant interpretation of the above results pertains to the equation of state (EoS) of the matter. If we set the integration constant $C_2 = 0$, the EoS of the matter ($\omega_m$) becomes:
\begin{eqnarray}
    \omega_m = \frac{\widetilde{p}_0}{\widetilde{\rho}_0} = \frac{M_0}{(4-5M_0)}.
\end{eqnarray}
We will soon see that the equation of state for the scalar field forces us to choose $M_0$ near its maximum value. It was shown in Ref.~\cite{Joshi1, Joshi2}
that $M_0$ can have a maximum value of $4/5$. In general, we will use the limits
\begin{eqnarray}
0 < M_0 \le \frac45\,.
\label{mlim}
\end{eqnarray}
If we have to interpret the scalar field as a remnant of the dark energy-like component in the late equilibrium state of the field-fluid system then we will require a high value of $M_0$. 

We can now calculate energy density and pressure corresponding to the scalar field. In our context the expressions for the energy density and the radial and tangential pressures are given by
\begin{eqnarray}
\rho_{\varphi} &=&  \frac{1}{2} \epsilon e^{- 2\beta} {\varphi^{\prime}}^2 + V(\varphi), \label{rhophidef}\\
    p^{(r)}_{\varphi} &=& \frac{1}{2} \epsilon e^{- 2\beta} {\varphi^{\prime}}^2 - V(\varphi), \label{prphidef} \\ 
    p^{(\vartheta)}_{\varphi} &=&  p^{(\phi)}_{\varphi} = -\frac{1}{2} \epsilon e^{- 2\beta} {\varphi^{\prime}}^2 - V(\varphi) = - \rho_{\varphi}. \label{ptanphidef}
\end{eqnarray}
Using Eq.~\eqref{phisoln3} and ~\eqref{v+fsoln3} with $\epsilon = -1$, it is now straightforward to obtain
\begin{eqnarray}
\rho_{\varphi} &=&\frac{M_0^3 }{16 (1 - M_0)^2}  \frac{1}{r^2} + C_2,\\
    p^{(r)}_{\varphi} &=& - \frac{M_0^2 (4 - 3M_0)}{16 (1 - M_0)^2}  \frac{1}{r^2} - C_2, \\
    p^{(\vartheta)}_{\varphi} &=&  p^{(\phi)}_{\varphi} = -\frac{M_0^3}{16 (1 - M_0)^2} \frac{1}{r^2} - C_2.
\end{eqnarray}
If we assume the value of the integration constant $C_2$ as zero, then the equation of state parameter for the scalar field ($\omega_{\varphi}$) which is defined as follows:
\begin{eqnarray}
    \omega_{\varphi} = \frac{p^{(r)}_{\varphi} + p^{(\vartheta)}_{\varphi} + p^{(\phi)}_{\varphi}}{3\rho_{\varphi}} = - \frac{4 - M_0}{3M_0}, \label{omegaphidef}
\end{eqnarray}
becomes a constant. Consequently the value of $\omega_{\varphi}$ is negative for the allowed range of values for $M_0$.   One must note here that scalar field sector has anisotropic pressure in spherically symmetric spacetimes. In FLRW spacetime the scalar field has isotropic pressure and has one parameter specifying the equation of state. In JMN-1 we actually have two different kinds of EoS for the scalar field, one corresponding to the radial pressure and the other corresponding with the angular components of pressure. For matching the equation of state with the conventional value it has in the cosmological sector, the above definition is used.  

We can see that the EoS parameter $\omega_{\varphi}$, which is a function of $M_0$, is a monotonously decreasing function of $M_0$. The maximum value of $\omega_{\varphi}$ is $-4/3$ ($\sim -1.34$)  which corresponds to $M_0 = 4/5$, and it steadily diverges to the value $- \infty $ as we decrease $M_0$ to it's minimum allowed value $0$. This shows that the scalar sector always acts like a dark energy component and the value of $\omega_\varphi$ is much less than that required from the cosmological sector. The present observational data requires the equation of state for the dark energy sector to be around $-1.03$ \cite{Escamilla:2023oce}.

As previously mentioned, the matter component of the composite fluid plays the role of dark matter. Since dark matter is typically considered a pressureless fluid, $\omega_m$ should tend to zero, implying $M_0 \to 0$. Given that the Schwarzschild mass is $M_{TOT} = \frac{M_0 R_b}{2}$, reducing the value of $M_0$ for a fixed Schwarzschild mass automatically increases the boundary radius ($R_b$) of the spherical overdense region. Therefore, the existence of dust-like dark matter implies a large value of $R_b$. This is compatible with our model of overdense regions, where we consider a composite fluid consisting of dark matter-like matter and a scalar field representing dark energy, and we know dark energy can have a non-zero effect on the structure formation of dark matter beyond a certain cosmological scale.

The crucial question is related to the the two EoS, $\omega_m$ and $\omega_\varphi$. We see both are functions of $M_0$, where $M_0$ is a parameter which defines the spacetime. For small values of $M_0$, near zero, the matter EoS can give a pressure-less fluid but the phantom EoS becomes a large negative number. On the other hand if we want to fix $\omega_\varphi$ close to $-1$, the value of $\omega_m$ becomes larger. There is no particular value for $M_0$ for which $p_m \sim 0$ and $p_\varphi \sim -\rho_\varphi$. Consequently it is difficult to assume that the present solution contains two components which act as the dark matter and dark energy as in the cosmological sector. It is seen that with proper choice of $M_0$ one can make $\omega_m$ near to zero. In that case the dark energy component is too negative if we assume the present values of the EoS remain the same in the cosmological length scales.  One of the ways in which this problem can be tackled is related to the non-constant nature of the EoS of the scalar field in the collapsing spacetime. The point is explained in the following paragraphs.

Assuming that the general form of the potential of the scalar field remains the same in both the cosmological and the structure formation scale and is given by
\begin{eqnarray}
V(\varphi)=V_0 \, e^{-\kappa \varphi}\,,
\label{scpot}
\end{eqnarray}
where $\kappa$ and $V_0$ depends upon the context, we can explain the origin of the two different kinds of EoS of the scalar field. Before we proceed with our discussion of the EoS we want to point out that the above form of the potential is remarkably similar to the form of the standard quintessence potential (or the phantom version of it) as used in Ref.~ \cite{Saha1, Saha2}. The result is interesting because till now we have not discussed the form of the quintessence potential nor have we discussed the cosmological evolution of the phantom field in FLRW spacetime. The form of the potential originating from the assumption that the phantom field is cohabiting with matter in a JMN-1 spacetime, yields the above form of the potential (setting $C_2$=0). It appears that the same scalar field can produce both acceleration of the background spacetime and seed a JMN-1 spacetime with matter, in a much smaller length scale compared to the cosmological scale,  at the expense of resetting the two constants $V_0$ and $\kappa$. The exact physics of this resetting of the potential parameters depend upon our knowledge about the dynamics which produces a static spacetime from an initially collapsing time dependent spacetime. At present our knowledge about such a transformation is not adequate.

The accelerating FLRW background forces the scalar field EoS to change and ultimately the EoS will tend towards $-1$. On the other hand the overdense region which detached from the background expansion will in general have a different kind of running of the scalar field EoS. Ultimately, when the collapsing system attains an equilibrium, the EoS of the scalar field in the overdense region need not be the same as the equation of state of the scalar field which is responsible for the accelerated expansion of the background universe. 

The scalar field, which is the remnant of the dark energy candidate in the detached overdense spacetime, becomes zero at $r=R_b$ and is only present inside a spherical region with dark matter. This is a perfect example of clustered dark energy, where the dark energy component do follow the dark matter sector in a gravitational collapse.

\subsubsection{Non-minimally coupled scalar field solutions for JMN-1 spacetimes}

Whether we have non-minimal interaction between the two components or they are noninteracting, in both the cases the scalar field is of the phantom type as specified in the previous discussion. If we introduce an interaction energy density arising from the algebraic type non-minimal coupling between a scalar field and matter, it takes the form \cite{Saha2, Tamanini1, Tamanini2, Tamanini3}:
\begin{eqnarray}
    \rho_{\text{int}} = \gamma \rho^{a}_m e^{-\xi \varphi} \equiv f(\rho_m, \varphi), \label{f}
\end{eqnarray}
then  the interacting pressure is defined as
\begin{eqnarray}
    p_{\text{int}} = n \frac{\partial f(n, \varphi)}{\partial n} -f(n, \varphi).
\end{eqnarray}
\begin{figure*}[]
\centering
\subfigure[]
{\includegraphics[width=86mm]{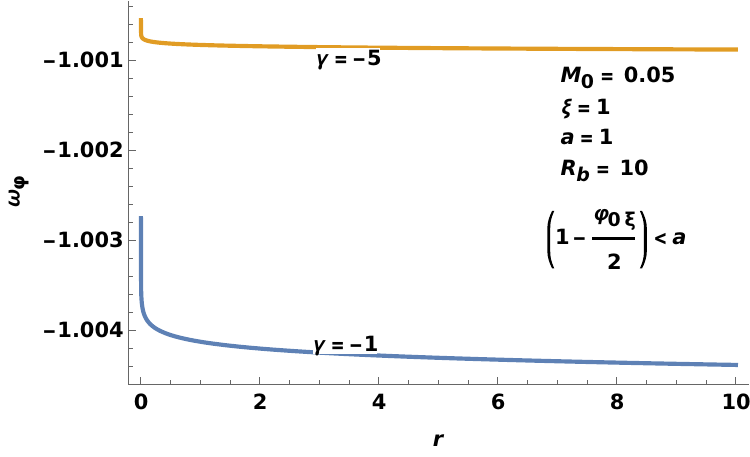}\label{EoSvsradiusa1}}
\hspace{0.5cm}
\subfigure[]
{\includegraphics[width=86mm]{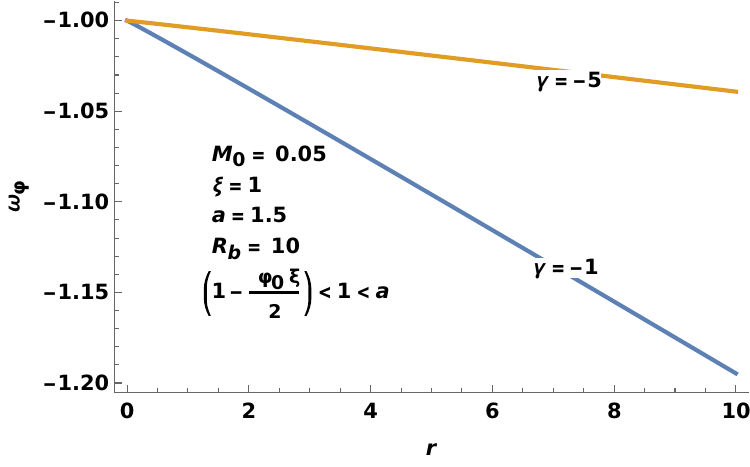}\label{EoSvsradiusa2}}
\hspace{0.5cm}
\subfigure[]
{\includegraphics[width=86mm]{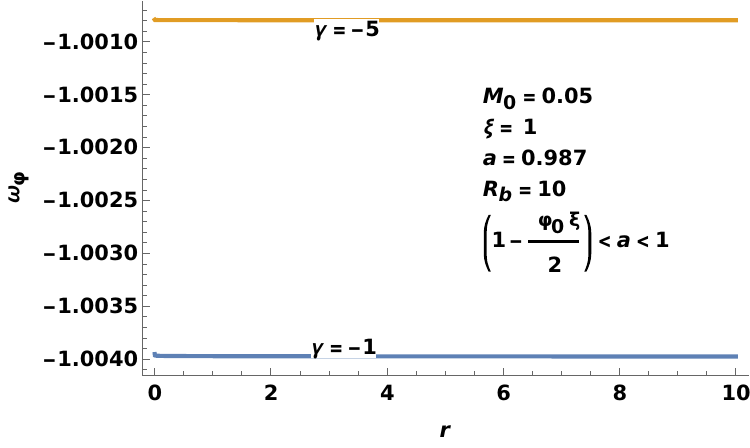}\label{EoSvsradiusa0p987}}
\hspace{0.5cm}
\subfigure[]
{\includegraphics[width=86mm]{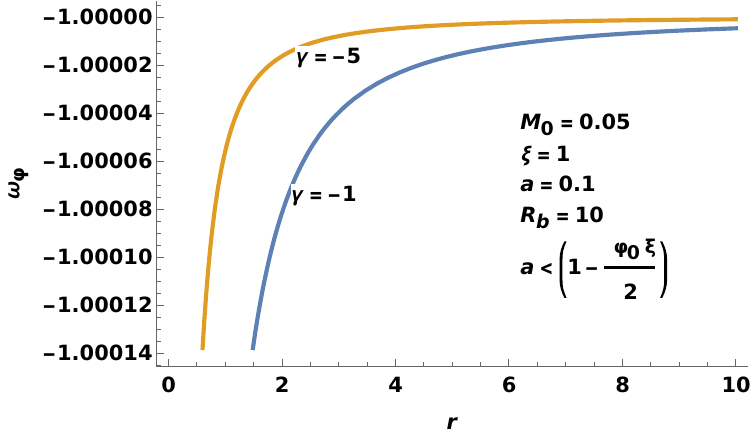}\label{EoSvsradiusa0p1}}
 \caption{The figures show how the equation of state of the scalar field ($\omega_\varphi$) varies with radial distance $r$ for different values of $a$, with the total Schwarzschild mass of the over-dense region being $M_S = 0.25$. The two curves, highlighted in blue and brown, illustrate the behavior of $\omega_\varphi$ for two different values of $\gamma$.}\label{omegaphi}
\end{figure*}
Here we have assumed that the matter energy density is a function of $n$, which is the particle number density. Here $a$ is a dimensionless real constant and $\xi$ is a dimensional constant. To calculate the interaction pressure, the above relation must be supplemented by a relation between $n$ and $\rho_m$. For simplicity, we assume the following relation:
\begin{eqnarray}
    \rho_m = k \hspace{0.5mm} n^{b},
\end{eqnarray}
where $k$ is a constant having appropriate dimension. The dimension of $k$ depends on the value of the real, dimensionless constant $b$. 
This relation between $n$ and $\rho_m$ implies
\begin{eqnarray}
    p_{\text{int}} &=& n \frac{\partial f(n,  \varphi)}{\partial \rho_m} \frac{\partial \rho_m}{\partial n} -f(n,  \varphi) \nonumber\\
    &=& n \left(\gamma a \rho^{a -1}_m e^{-\xi \varphi}\right) \left( k b n^{b-1}\right) - \gamma \rho^{a}_m e^{-\xi \varphi} \nonumber \\
    &=& (ab - 1) \gamma \rho^{a}_m e^{-\xi \varphi} \nonumber \\
    &=& (ab - 1) f(\rho_m, \varphi).
\end{eqnarray}
So, using Eq.~\eqref{phisoln3} and Eq.~\eqref{f}, the interacting energy density and interaction pressure becomes
\begin{eqnarray}
   \rho_{\text{int}}(r) &=& {\gamma} \left(\frac{\widetilde{\rho}_0}{r^2} - C_2 \right)^a \left(\frac{r}{R_b}\right)^{-\xi \varphi_0}, \label{rhointjmn1}\\
   p_{\text{int}}(r) &=& {\gamma} (ab-1)\left(\frac{\widetilde{\rho}_0}{r^2} - C_2 \right)^a \left(\frac{r}{R_b}\right)^{-\xi \varphi_0}. \label{pintjmn1}
\end{eqnarray}
\begin{figure*}[]
\centering
\subfigure[]
{\includegraphics[width=86mm]{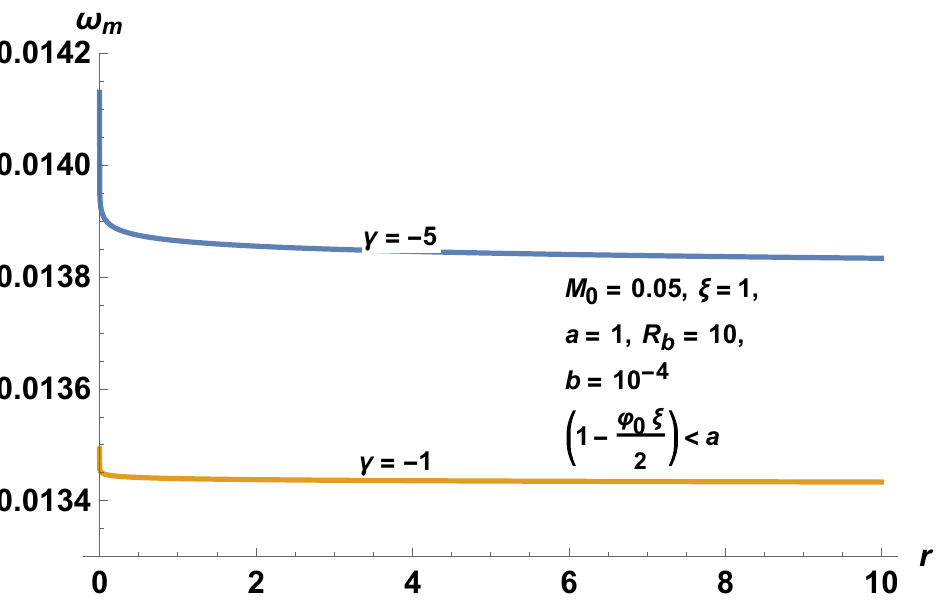}\label{EoSmvsradiusa1}}
\hspace{0.5cm}
\subfigure[]
{\includegraphics[width=86mm]{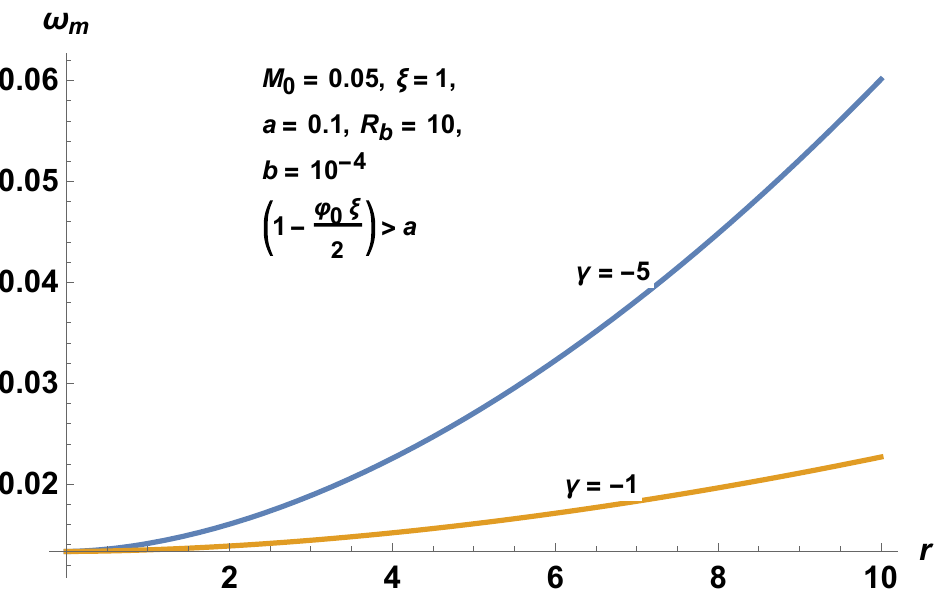}\label{EoSmvsradiusa0p1}}
 \caption{The figures illustrate how the equation of state of the scalar field ($\omega_m$) varies with radial distance $r$ for different values of $a$, given a total Schwarzschild mass of the over-dense region $M_S = 0.25$. The two curves, highlighted in blue and brown, depict the behavior of $\omega_m$ for two different values of $\gamma$.}\label{omegaphi}
\end{figure*}
In the present case, it can be observed that if $a=b=1$ then the interaction pressure vanishes although the interaction energy density remains nonzero. Here $\varphi_0$ and $\widetilde{\rho}_0$ were defined in the previous subsection.

The general solution for energy density, pressure and potential function can be obtained from Eqs.~\eqref{ab}, ~\eqref{cd} and \eqref{v+fsolngen} as the following:
\begin{align}
    &p_m(r) + p_{\text{int}}(r) = V(\varphi) + \frac{M^2_0}{8(1-M_0)}  \frac{1}{r^2}, \label{int_psoln_jmn1} \\
    &V(\varphi(r)) + f = C_2 + \frac{M^2_0 (2-M_0)}{16(1-M_0)^2} \frac{1}{r^2}, \label{int_v+fsoln_jmn1} \\
    &\rho_m(r) = - C_2 + \frac{M_0(4-3M_0)(4-5M_0)}{16(1-M_0)^2} \frac{1}{r^2}. \label{int_rhosoln_jmn1} 
\end{align}
Using Eq.~\eqref{int_v+fsoln_jmn1}, ~\eqref{f} and ~\eqref{phisoln3}, we get the $\varphi$-dependence of the potential 
\begin{eqnarray}
    V(\varphi) = C_2 + \widetilde{V}_0 e^{-2\varphi/\varphi_0} - \gamma\left(\frac{\widetilde{\rho}_0}{R^2_b} e^{-{2\varphi}/{\varphi_0}}  - C_2 \right)^a  e^{-\xi \varphi}. \nonumber\\
    \label{intV}
\end{eqnarray}
Using Eq.~\eqref{int_psoln_jmn1} and substituting the expressions for $p_{\text{int}}$ and $V(r) + f$ from Eqs.~\eqref{pintjmn1} and \eqref{int_v+fsoln_jmn1}, we get
\begin{align}
    p_m(r) = C_2 + \frac{\widetilde{p}_0}{r^2} - ab {\gamma} \left(\frac{\widetilde{\rho}_0}{r^2} - C_2 \right)^a  \left(\frac{r}{R_b}\right)^{-\xi \varphi_0}.
    \label{intpm}
\end{align}
However, in the interacting case, $\rho_m(r)$ retains its previous isothermal form:
\begin{eqnarray}
    \rho_m(r) = \frac{\widetilde{\rho}_0}{r^2} - C_2. \label{intrhom}
\end{eqnarray}

Using these expressions for the energy density, pressure and potential function, we can calculate the EoS parameter $\omega_{\varphi}$ for the non-minimally coupled case. Using the definitions from Eqs.~\eqref{rhophidef}, \eqref{prphidef} and \eqref{ptanphidef} and the expression for $\varphi(r)$ from Eq.~\eqref{phisoln3} (which is valid for the non-minimally coupled case also), and the potential function from Eq.~\eqref{intV}, we have the following results:

\begin{align}
\rho_{\varphi} &=  \frac{\widetilde{p}}{r^2} - {\gamma} \left(\frac{\widetilde{\rho}_0}{r^2} - C_2 \right)^a \left(\frac{r}{R_b}\right)^{-\xi \varphi_0} + C_2, \label{rhophijmn1int}\\
    p^{(r)}_{\varphi} &= -  \frac{\widetilde{p}_0}{r^2} + {\gamma} \left(\frac{\widetilde{\rho}_0}{r^2} - C_2 \right)^a \left(\frac{r}{R_b}\right)^{-\xi \varphi_0} - C_2, \label{prphijmn1int}\\
    p^{(\vartheta)}_{\varphi} &= p^{(\phi)}_{\varphi} = - \frac{\widetilde{p}_{}}{r^2} + {\gamma} \left(\frac{\widetilde{\rho}_0}{r^2} - C_2 \right)^a \left(\frac{r}{R_b}\right)^{-\xi \varphi_0} - C_2, \label{ptanphijmn1int}
\end{align}
where $\widetilde{p}_0$ is defined in Eq.~\eqref{ptild} and
\begin{eqnarray}
    \widetilde{p}_{} = \frac{M_0^3}{16 (1 - M_0)^2}.
\end{eqnarray}
Using the definition in Eq.~\eqref{omegaphidef} and the expressions for $\rho_{\varphi}$, $p^{(r)}_{\varphi}$, $p^{(\vartheta)}_{\varphi}$ and $p^{(\phi)}_{\varphi}$ from Eqs.~\eqref{rhophijmn1int}, \eqref{prphijmn1int} and \eqref{ptanphijmn1int} we can now obtain 
\begin{eqnarray}
    \omega_{\varphi} = -\frac{\widetilde{p}_0 + 2\widetilde{p} - 3\gamma {\widetilde{\rho}_0}^a {R_b}^{\xi \varphi_0}r^{2 - 2a - \xi \varphi_0}}{3\widetilde{p}  -  3\gamma {\widetilde{\rho}_0}^a {R_b}^{\xi \varphi_0} r^{2 - 2a - \xi \varphi_0}},
\end{eqnarray}

where we have assumed $C_2 = 0$. In terms of the spacetime parameter $M_0$, the above expression for $\omega_\varphi$ can be written as:
\begin{eqnarray}
   \omega_{\varphi}= \frac{\frac{(M_0-4) M_0^2}{(M_0-1)^2}+48 \gamma  \Psi \left(\frac{R_b}{r}\right)^{\frac{M_0 \xi }{2(1- M_0)}} r^{2(1-a)}}{\frac{3  M_0^3}{ (M_0-1)^2}-48 \gamma  \Psi \left(\frac{R_b}{r}\right)^{\frac{M_0 \xi }{2(1- M_0)}} r^{2(1-a)}}, \label{omega_0}
\end{eqnarray}
where $\Psi=\left\{\frac{M_0 (3 M_0-4) (5 M_0-4)}{16(M_0-1)^2}\right\}^a$. This expression of $\omega_\varphi$ describes its behavior throughout the over-dense region (i.e., $\forall r\in (0,R_b]$) modeled by JMN-1 spacetime. It can be verified from the expression of $\Psi$ that for $M_0=0.8$, $\forall r\in (0,R_b]$, $\omega_\varphi$ remains constant: $\omega_\varphi=-\frac{4}{3}$. However, for $M_0<0.8$ and $a>\left(1-\frac{\varphi_0 \xi}{2}\right)$, $\omega_\varphi$ takes the following values near the center and the boundary:
\begin{eqnarray}
\omega_\varphi=\begin{cases}
-1, & \text{when } r \to 0\\
\\
\frac{\frac{(M_0-4) M_0^2}{(M_0-1)^2}+48 \gamma  \Psi  R_b^{2(1-a)}}{\frac{3 M_0^3}{ (M_0-1)^2}-48 \gamma  \Psi  R_b^{2(1-a)}}, & \text{when } r \sim R_b.
\end{cases}
\label{omega2}
\end{eqnarray}
For $a=1$, which satisfies the condition $a>\left(1-\frac{\varphi_0 \xi}{2}\right)$ for $\xi>0$, the expression for $\omega_\varphi$ near $r=R_b$ simplifies to:
\begin{eqnarray}
    \omega_\varphi=-1+\frac{4(1-M_0)M_0}{3\left[\gamma(4-3M_0)(4-5M_0) - M_0^2\right]}.
\end{eqnarray}

Now, if we consider small values of $M_0$, we can write $\omega_\varphi \sim -1+\frac{M_0}{12\gamma}$ near $r=R_b$ for $a=1$. Therefore, for $a=1$ and $\gamma<0$ (or $\gamma>0$), the expression of $\omega_\varphi$ at $r=R_b$ suggests that $\omega_\varphi$ tends to increase (or decrease) towards (from) $-1$ as the value of $M_0$ decreases or the absolute value of $\gamma$ increases. Fig.~\ref{EoSvsradiusa1} illustrates the behavior of $\omega_\varphi$ as a function of radial distance $r$, with parameters $a=1$, $M_0=0.05$, $\xi=1$, and $R_b=10$. In this figure, it can be seen that near the center, $\omega_\varphi \sim -1$. Although $\omega_\varphi$ decreases with radial distance, it remains close to $-1$ throughout the over-dense region, with a higher absolute value of $\gamma$ causing it to stay even closer to $-1$.

For $a>1$ and for given values of $M_0$ and $\gamma$, Eq.~(\ref{omega_0}) implies that near the boundary, $\omega_\varphi$ tends to $-\frac{(4-M_0)}{3M_0}$ for large values of $R_b$, since the term $48 \gamma  \Psi  R_b^{2(1-a)}$ becomes negligible for large $R_b$. As discussed earlier, for a given Schwarzschild mass of the overdense region, $M_0$ is inversely proportional to $R_b$. Therefore, a large $R_b$ implies small values of $M_0$. In Fig.~\ref{EoSvsradiusa2}, we illustrate the same concept discussed above. In this figure, we consider $a=1.5$ while keeping the values of all other parameters the same as in Figs.~\ref{EoSvsradiusa1}, ~\ref{EoSvsradiusa0p987}, and ~\ref{EoSvsradiusa0p1}. It can be seen that, in this case, $\omega_\varphi$ is no longer close to $-1$ and steadily decreases towards the boundary. The rate of decrease in the value of $\omega_{\varphi}$ is greater for a lower absolute value of $\gamma$ as can be seen from the graph. Consequently, choosing a suitable negative value of $\gamma$, we can keep the value of $\omega_{\varphi}$ close to $-1$.

As $\Psi$ varies as $M_0^a$ and as $M_0$ is inversely proportional to $R_b$ for a fixed Schwarzschild mass, we have $\Psi \sim {R_b}^{-a}$. Considering this, we see that for $\left(1-\frac{\varphi_0 \xi}{2}\right)<a<\frac{2}{3}$, $\omega_\varphi$ tends to $-1$ at the boundary when the boundary radius $R_b$ becomes large. This analysis reveals that $\gamma$ and $R_b$ need to be large (i.e., $M_0$ needs to be small for a given Schwarzschild mass) with $\left(1-\frac{\varphi_0 \xi}{2}\right)<a<\frac{2}{3}$ to make $\omega_\varphi$ close to $-1$ throughout the over-dense region. We illustrate this nature of $\omega_\varphi$ in Fig.~\ref{EoSvsradiusa0p987} where we consider $a=0.987$. Here, we consider $M_0=0.05$ and $\xi=1$. Therefore, $\left(1-\frac{\varphi_0 \xi}{2}\right)=0.98684$, which implies $\left(1-\frac{\varphi_0 \xi}{2}\right)<a<1$. In all the sub-figures of Fig.~(\ref{omegaphi}), we consider fixed Schwarzschild mass $M_{S} = 2.5$.

Now, for $a<\left(1-\frac{\varphi_0 \xi}{2}\right)$, $\omega_\varphi$ takes the following values near the center and the boundary: 
\begin{eqnarray}
\omega_\varphi=\begin{cases}
-\frac{(4-M_0)}{3 M_0}, & \text{when } r \to 0\\
\\
\frac{\frac{(M_0-4) M_0^2}{(M_0-1)^2}+48 \gamma  \Psi  R_b^{2(1-a)}}{\frac{3 M_0^3}{ (M_0-1)^2}-48 \gamma  \Psi  R_b^{2(1-a)}}, & \text{when } r \sim R_b.
\end{cases}
\label{omega3}
\end{eqnarray}
This implies that near the center, $-\infty <\omega_\varphi \leq -\frac{4}{3}$ since $0<M_0\leq 0.8$. Since $a<\left(1-\frac{\varphi_0\xi}{2}\right)$ always implies $a<1~\forall~M_0\in(0,0.8),~ \xi>0$, near the boundary $\omega_\varphi$ tends to $-1$ for large values of $R_b$. Therefore, for sufficiently large values of $\gamma$ and $R_b$, the scenario $a<\left(1-\frac{\varphi_0\xi}{2}\right)$ always results in $\omega_\varphi$ being close to $-1$ in a large portion of the overdense region. This behavior of $\omega_\varphi$ is illustrated in Fig.~\ref{EoSvsradiusa0p1}, where $\omega_\varphi$ approaches a value distinctly smaller than $-1$ (not shown in the graph because of the chosen range of the $\omega_\varphi$ axis) as the radial coordinate $r$ approaches the central singularity at $r = 0$, then sharply increases towards $-1$ as $r$ increases. Higher values of $|\gamma|$ result in a more pronounced decrease in $\omega_\varphi$. It can be noticed that in all the figures we consider negative values of $\gamma$. This is because, for positive values of $\gamma$, $\omega_\varphi$ has a pole at $$r = \left\{\frac{M_0^3 {\Psi}^{-1} R_b^{-\frac{M_0 \xi}{2(1-M_0)}}}{16\gamma (1-M_0)^2}\right\}^{\frac{2(1-M_0)}{4-4a(1-M_0)-M_0(4+\xi)}},$$ which is not a physically viable scenario. It can be verified that the pole vanishes when $\gamma<0$. Therefore, we will consider only negative values of $\gamma$ in our further analysis.

\begin{widetext}
\begin{table}
\captionsetup{width=\textwidth}
\renewcommand{\arraystretch}{1.4} 
\setlength{\tabcolsep}{3pt}      
\begin{tabular}
{|p{3.5cm}|p{3.5cm}|p{3.5cm}|p{3cm}|p{4cm}|}
\hline
Parameter's range & $\omega_\varphi$ at $r\to 0$ & $\omega_\varphi$ at $r\to R_b$ & $\omega_m$ at $r\to 0$ & $\omega_m$ at $r\to R_b$ \\
\hline
$M_0<0.8$, $\gamma<0$, and $a>\left(1-\frac{\varphi_0 \xi}{2}\right)$  
& -1 
& $\frac{\frac{(M_0-4) M_0^2}{(M_0-1)^2}+48 \gamma  \Psi  R_b^{2(1-a)}}{\frac{3 M_0^3}{ (M_0-1)^2}-48 \gamma  \Psi  R_b^{2(1-a)}}$ 
& $\infty$ 
& $\frac{M_0}{4-5M_0}-ab\gamma\Psi^{\frac{a-1}{a}}R_b^{2(1-a)}$ \\
\cline{3-3}
& 
& $\sim -1+\frac{M_0}{12\gamma}$, for $M_0 \ll 1$ and $a=1$. 
& 
& \\
\cline{3-3}                 
& 
& $\sim -\frac{(4-M_0)}{3M_0}$, for $M_0 \ll 1$ and $a>1$.
& 
& \\
\cline{3-3}
& 
& $\sim -1$ for $\left(1-\frac{\varphi_0 \xi}{2}\right)<a<\frac{2}{3}$, $M_0 \ll 1$, and for large value of $|\gamma|$.
& 
& \\
\hline
$M_0<0.8$, $\gamma<0$, and $a<\left(1-\frac{\varphi_0 \xi}{2}\right)$ 
& $-\frac{(4-M_0)}{3 M_0}$ 
& $\frac{\frac{(M_0-4) M_0^2}{(M_0-1)^2}+48 \gamma  \Psi  R_b^{2(1-a)}}{\frac{3 M_0^3}{ (M_0-1)^2}-48 \gamma  \Psi  R_b^{2(1-a)}}$ 
& $\frac{M_0}{4-5M_0}$ 
&$\frac{M_0}{4-5M_0}-ab\gamma\Psi^{\frac{a-1}{a}}R_b^{2(1-a)}$ \\
\cline{3-3}
& 
& $\sim -1$ for $M_0 \ll 1$, and for large value of $|\gamma|$. 
& 
& \\
\hline
\end{tabular}
\caption{Nature of the equation of state parameters, $\omega_\varphi$ and $\omega_m$, of a non-minimally coupled scalar field and matter, respectively, across different parameter spaces in the JMN-1 spacetime.}
\label{table1}
\end{table}
\end{widetext}
Next, we will study the equation of state of the matter sector of the composite fluid. From Eqs.~\eqref{intpm} and \eqref{intrhom} it is straightforward to calculate the EoS parameter for the matter sector, which gives \begin{eqnarray}
 \omega_m = \frac{\widetilde{p}_0}{\widetilde{\rho}_0} - \left(ab \gamma {\widetilde{\rho}_0}^{a-1} R^{\xi \varphi_0}_b \right) r^{2 - 2a - \xi \varphi_0},
\end{eqnarray}
where we have assumed $C_2 = 0$ like before. The above expression of  $\omega_m$ can be written in the following form in terms of spacetime parameters:
\begin{eqnarray}
    \omega_m~=~\frac{M_0}{4-5M_0}-ab\gamma\Psi^{\frac{a-1}{a}}r^{2(1-a)}\left(\frac{R_b}{r}\right)^{\xi\varphi_0}.
\end{eqnarray}
Similar to the previous analysis for $\omega_\varphi$, here also we can study the nature of $\omega_m$ for $a>\left(1-\frac{\varphi_0\xi}{2}\right)$ and $a<\left(1-\frac{\varphi_0\xi}{2}\right)$.

For $a>\left(1-\frac{\varphi_0\xi}{2}\right)$, it can be verified that near the center and the boundary $\omega_m$ have the following form:
\begin{eqnarray}
\omega_m=\begin{cases}
\infty, & \text{when } r \to 0\\
\\
 \frac{M_0}{4-5M_0}-ab\gamma\Psi^{\frac{a-1}{a}}R_b^{2(1-a)}, & \text{when } r \sim R_b,
\end{cases}
\label{omegam1}
\end{eqnarray}
whereas for $a<\left(1-\frac{\varphi_0\xi}{2}\right)$ we can write: 
\begin{eqnarray}
\omega_m=\begin{cases}
\frac{M_0}{4-5M_0}, & \text{when } r \to 0\\
\\
 \frac{M_0}{4-5M_0}-ab\gamma\Psi^{\frac{a-1}{a}}R_b^{2(1-a)}, & \text{when } r \sim R_b.
\end{cases}
\label{omegam2}
\end{eqnarray}
Since an infinitely large value for the equation of state parameter of the matter component is not physically possible, we do not consider any results for $a > \left(1 - \frac{\varphi_0 \xi}{2}\right)$. Although we obtained physically viable results for $\omega_\varphi$ in that range, we exclude them from further consideration.
For $a < \left(1 - \frac{\varphi_0 \xi}{2}\right)$, $\omega_m$ has a finite value near the center, specifically $\frac{M_0}{4 - 5M_0}$. This value approaches zero for sufficiently small values of $M_0$, which is ideal for the matter component of the composite fluid. It should be noted that $M_0<<0.8$, since $\omega_m$ diverges at $M_0=0.8$. 
Near the boundary, to ensure $\omega_m$ remains close to its value at the center, we need to consider sufficiently small values of $b$. We cannot assume $\gamma$ to be close to zero because this would make $\omega_\varphi$ deviate from $-1$, which is not ideal for the scalar field representing dark energy. Fig.~\ref{EoSmvsradiusa1} shows the behavior of $\omega_m$ as a function of $r$ for $a = 1$, $M_0 = 0.05$, $R_b = 10$, $b = 10^{-4}$, and $\xi = 1$. It is observed that $\omega_m$ tends to infinity near the center for both $\gamma = -5$ and $\gamma = -1$. In contrast, for $a = 0.1$, which satisfies the condition $a < \left(1 - \frac{\varphi_0 \xi}{2}\right)$, Fig.~\ref{EoSmvsradiusa0p1} shows that $\omega_m$ remains finite throughout the over-dense region and stays close to zero. In Table \ref{table1}, we present the behavior of $\omega_\varphi$ and $\omega_m$ across various parameter spaces.

From the above analysis, it can be concluded that to make the matter part and scalar field part of the composite fluid behave like dark matter and dark energy, respectively, we need to consider $a < \left(1 - \frac{\varphi_0 \xi}{2}\right)$, large negative values of $\gamma$, small values $M_0$ and $b$. From Eq.~(\ref{rhointjmn1}) and Eq.~(\ref{pintjmn1}), it can be observed that a large absolute value of $\gamma$ results in a significant interaction-energy density ($\rho_{int}$) and pressure ($p_{int}$). It also can be verified that $\rho_{int}$ and $p_{int}$ diverge near the central singular point for both ranges of $a$, i.e., $a < \left(1 - \frac{\varphi_0 \xi}{2}\right)$ and $a > \left(1 - \frac{\varphi_0 \xi}{2}\right)$.  With these constraints in the parameter space, we can model JMN-1 spacetime seeded by dark matter non-minimally coupled with a dark energy-like scalar field. 

The components seeding the JMN-1 spacetime can be interpreted as remnants of the original dark matter and dark energy components after gravitational collapse and virialization. Virialization modifies the EoS of the dark sector components.
Although Eq.~\eqref{omega3} shows that $\omega_\varphi$ approaches a value distinctly smaller than $-1$ as $r \to 0$, it can also be seen that $\omega_\varphi$ sharply increases to values near $-1$ as $r$ increases. 
The existence of a spacetime singularity at the center might explain this change in the equation of state of dark energy. To exist in the ultra-high curvature region with matter with an almost zero equation of state, the EoS of dark energy must increase, but it does not exceed $-\frac{1}{3}$, thus violating the strong energy condition.

We conclude this section with a discussion on the energy conditions for the JMN-1 spacetime. From Eqs.~\eqref{f0soln2}, ~\eqref{f1soln2} and ~\eqref{f2soln2}, we have
\begin{align}
    \rho &= \frac{M_0}{r^2}, \label{tot_rho_p_jmn-1} \\
    p^{(r)} = 0, \,\ 
    p^{(\vartheta)} &= p^{(\phi)} =  \frac{M_0^2}{4(1-M_0) r^2}. \nonumber 
\end{align}
Here $\rho$, $p^{(r)}$, $p^{(\vartheta)}$ and $p^{(\phi)}$ are the total energy density, total pressure along the radial direction and total transverse stress of the constituents in JMN-1 spacetime.  As the parameter $M_0$ is always positive and $0 < M_0 \leq \frac{4}{5}$, we have $\rho > 0$, $p^{(\vartheta)} > 0$, $p^{(\phi)} > 0$,  and consequently $\rho + p^{(i)} > 0, \forall i$; where $i$ represents any of the pressure components $p^{(r)}$, $p^{(\vartheta)}$ or $p^{(\phi)}$. Consequently we also have $\rho + p^{(r)} + p^{(\vartheta)} + p^{(\phi)} > 0$. From Eq.~\eqref{tot_rho_p_jmn-1}, we can calculate, that the condition $\rho \geq \left|p^{(i)}\right|$ translates to $M_0 \leq \frac{4}{5}$. So all the energy conditions, viz., weak, strong, null and dominant; are satisfied in JMN-1 spacetime.


\subsubsection{Minimally coupled scalar field solutions for JMN-2 spacetimes}

Comparing the line element of JMN-2 (Eq.~(\ref{JMN2metric})) spacetime with the line element of static spherically symmetric spacetime given in Eq.~(\ref{metric1}), we get:
\begin{eqnarray}
    \alpha(r) &=& - \frac{1}{2} \ln\left[16 \lambda^2 (2-\lambda^2) \right] \nonumber\\&&+ \ln\left[(1+\lambda)^2 \left(\frac{r}{R_b} \right)^{1-\lambda} - (1-\lambda)^2 \left(\frac{r}{R_b} \right)^{1+\lambda} \right],\nonumber \\
    \beta(r) &=& \frac{1}{2} \ln\left[2-\lambda^2 \right],
\end{eqnarray}
where $0\leq \lambda < 1$.

The energy-momentum tensor components $T^0_0$, $T^1_1$ and $T^2_2$ for JMN-2 spacetime are given by
\begin{eqnarray}
    T^0_0 &=& - \left(\frac{1-\lambda^2}{2-\lambda^2}\right)\frac{1}{r^2}, \label{f0soln22} \\
    T^i_j &=& \frac{1}{(2-\lambda^2)}\frac{1}{r^2}\left[\frac{(1-\lambda)^2 A - (1+\lambda)^2 B r^{2\lambda}}{A - B r^{2\lambda}}\right]\delta^i_j.\nonumber\\
    \label{f1soln22}.
\end{eqnarray}
Eq.~\eqref{f1soln22} indicates that the composite fluid seeding the JMN-2 spacetime is isotropic in nature. Consequently, as previously stated:
\begin{eqnarray}
    \varphi^{\prime} (r) = 0 \implies \varphi = \text{constant} \equiv \varphi_0,
\end{eqnarray}
where we use Eq.~(\ref{2e3}). From Eq.~(\ref{phi}) it is seen that this constant scalar field can be a canonical scalar field or a phantom like scalar field, JMN-2 spacetime cannot differentiate between these two kinds of scalar fields.

As in this spacetime we have $T^1_1 = T^2_2$, from Eq.~\eqref{v+f} we obtain a solution for the potential function $V$ after substituting $f$ with zero
\begin{eqnarray}
    V(r) = C_2 = \text{constant} \equiv V_0.
\end{eqnarray}
\begin{figure*}[]
\centering
\subfigure[]
{\includegraphics[width=86mm]{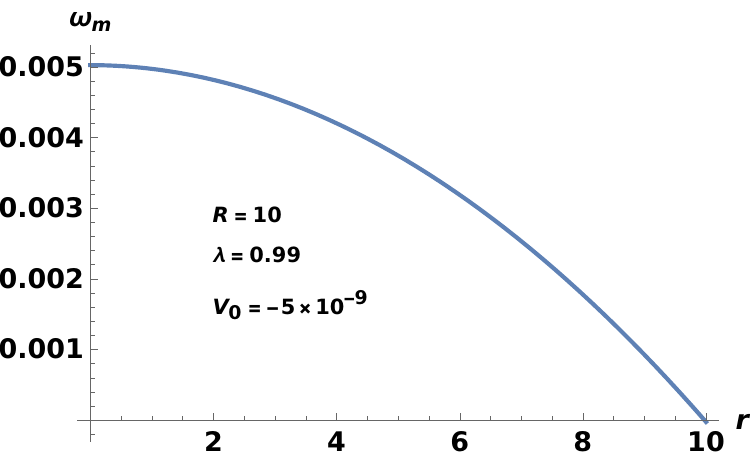}\label{EoS_vs_r_jmn2_minimal_1}}
\hspace{0.5cm}
\subfigure[]
{\includegraphics[width=86mm]{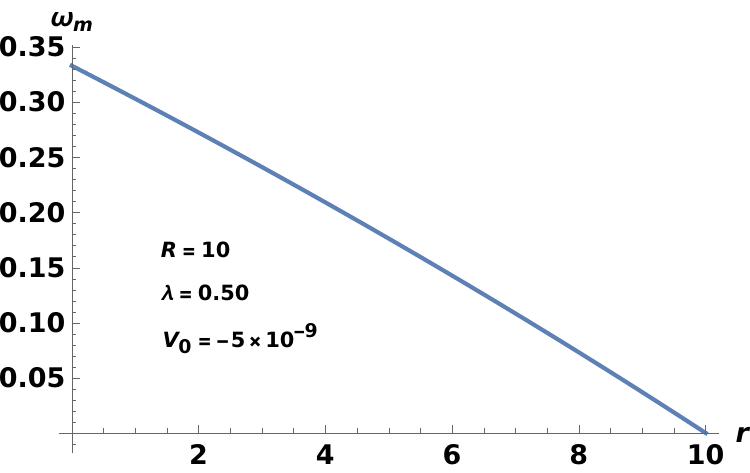}\label{EoS_vs_r_jmn2_minimal_2}}
 \caption{The figures show how the equation of state parameter of the matter sector ($\omega_m$) varies with radial distance $r$ in JMN-2 spacetime when the matter is minimally coupled with the scalar field. For comparison we have shown two different figures for two different values of $\lambda$. In both of these plots, the value of the constant $V_0$ has been kept very close to zero, as mentioned in the text. We can see that (Fig~(\ref{EoS_vs_r_jmn2_minimal_2}), as the value of the parameter $\lambda$ decreases, the curve becomes a straight line, showing the steady linear decrease of $\omega_m$ with the radial coordinate $r$.}\label{}
\end{figure*}
We can obtain the expressions for the energy density and pressure from the Eqs.~\eqref{ab} and ~\eqref{cd} after subtracting and adding them respectively:
\begin{align}
  \rho_m &= - V(\varphi) + \frac{T^2_2 - T^1_1}{2} - T^0_0, \\
    p_m &= V(\varphi) + \frac{T^1_1 + T^2_2}{2}.  
\end{align}

Using Eqs.~\eqref{f0soln22} and ~\eqref{f1soln22} the above expressions become
\begin{align}
    \rho_m &=  \left(\frac{1-\lambda^2}{2-\lambda^2}\right)\frac{1}{r^2} - V_0, \\
   p_m &= \frac{1}{(2-\lambda^2)}\frac{1}{r^2}\left[\frac{(1-\lambda)^2 A - (1+\lambda)^2 B r^{2\lambda}}{A - B r^{2\lambda}}\right] + V_0. 
\end{align}
From the expressions for $\rho_m$ and $p_m$ we can now calculate the EoS parameter for the matter sector
\begin{align}
    \omega_m = \frac{p_m}{\rho_m} = \frac{\Delta_1(\lambda, r) A - \Delta_2(\lambda, r) B r^{2 \lambda}}{A - B r^{2\lambda}}, \label{omegamjmn2minimal}
\end{align}
where $\Delta_1$ and $\Delta_2$ are defined as follows:
\begin{align}
    \Delta_1(\lambda, r) = \frac{(1-\lambda)^2 + V_0 (2-\lambda^2) r^2}{1-\lambda^2 - V_0 (2-\lambda^2) r^2}, \\
    \Delta_2(\lambda, r) = \frac{(1+\lambda)^2 + V_0 (2-\lambda^2) r^2}{1-\lambda^2 - V_0 (2-\lambda^2) r^2}.
\end{align}
We can now see the qualitative nature of the EoS for matter in the present scenario.

It is straightforward to check the limiting behaviour of $\omega_m$ as we approach from the boundary of the overdense region ($r = R_b$) towards the centre ($r =0$):
\begin{align}
\omega_m=\begin{cases}
\frac{1-\lambda}{1+\lambda}, \hspace{3cm} \text{when } r \to 0\\
\\
 \frac{1}{4\lambda}\left[(1+\lambda)^2 \Delta_1(\lambda, R_b) - (1-\lambda)^2 \Delta_2(\lambda, R_b) \right], \\  \hspace{38mm} \text{when } r \sim R_b.
\end{cases}
\label{omegamlimitjmn2}    
\end{align}
From the above limiting cases it can be seen that, if we want to model the matter component as pressure-less dark matter, then the value of $\lambda$ should be very close to $1$ to make the value of $\omega_m$ very small at the centre of the spherical region. It can also be understood from the expressions of $\Delta_1$ and $\Delta_2$, that, to avoid any zero in the denominator of $\omega_m$ (i.e., to avoid the divergence of $\omega_m$) we should consider only negative values of $V_0$.

From the expression of $\omega_m$ in Eq.~\eqref{omegamlimitjmn2} we can see that, the value of this quantity will be close to zero at the boundary of the spherical region, if  
\begin{align}
    (1+\lambda)^2 \Delta_1(\lambda, R_b) - (1-\lambda)^2 \Delta_2(\lambda, R_b) \approx 0,
\end{align}
or equivalently, $\Delta_1(\lambda, R_b) / \Delta_2(\lambda, R_b) = [(1-\lambda)/(1+\lambda)]^2$. From the expressions of $\Delta_1$ and $\Delta_2$ we have
\begin{align}
    \frac{\Delta_1(\lambda, R_b)}{\Delta_2(\lambda, R_b)} = \frac{(1+\lambda)^2 + V_0 (2-\lambda^2) R^2_b}{(1-\lambda)^2 + V_0 (2-\lambda^2) R^2_b}.
\end{align}
We can make the value of this expression arbitrarily close to the value of  $[(1-\lambda)/(1+\lambda)]^2$ if we assume
\begin{align}
   \left|V_0 R^2_b\right|  \ll (1-\lambda)^2,
\end{align}
for $\lambda \approx 1$. Consequently, unless we do not want the size of the spherical overdense region to be very small, we have to assume that the value of $V_0$ is very near to zero. If we plot the function $\omega_m$ from 
Eq.~\eqref{omegamjmn2minimal} for a very small negative value of $V_0$ and the value of $\lambda$ being very close to $1$, we can see (Fig~\ref{EoS_vs_r_jmn2_minimal_1} and ~\ref{EoS_vs_r_jmn2_minimal_2}) that $\omega_m$ varies from a small value at the centre and steadily decreases towards $0$ as we approach $R_b$. 

As we have a constant and homogeneous scalar field profile, the value of the EoS parameter corresponding to the scalar field sector is 
\begin{align}
    \omega_{\varphi} = -1.
\end{align}
This shows that the scalar field sector behaves like the cosmological constant inside the spherical region of our interest. 
So we can conclude that, to model the matter part as dark matter in JMN-2 spacetime, we need to keep the value of the parameter $\lambda$ very close to $1$ and $V_0$ near to $0$ respectively, i.e., $\lambda \approx 1$ and $|V_0| \approx 0$.



\subsubsection{Non-minimally coupled scalar field solutions for JMN-2 spacetimes}

We should remind the reader that the scalar field in our case can be either a canonical scalar field or a phantom like scalar field, non-minimally coupled components in JMN-2 spacetime cannot differentiate between these two kinds of scalar fields.
Using the Klein-Gordon equation for $\varphi (r)$, i.e., Eq.~\eqref{phieom}, and noting that in this case the scalar field solution is actually a constant, we have:
\begin{eqnarray}
    \frac{\partial V(\varphi)}{\partial \varphi} + \frac{\partial f(\rho_m, \varphi)}{\partial \varphi} = 0,
\end{eqnarray}
which has a general solution:
\begin{eqnarray}
    V(\varphi) + f(\rho_m, \varphi) = \widetilde{\zeta}(\rho_m) + \widetilde{C}_2, \label{v+fjmn2p0}
\end{eqnarray}
where $\widetilde{\zeta}(\rho_m)$ is an arbitrary function of the matter-energy density $\rho_m$, and $\widetilde{C}_2$ is an integration constant. The above equation shows that the effective interaction term $f(\rho_m, \varphi)$ in the present case cannot be written as a multiplicative interaction term as given in Eq.~(\ref{mulf}). Here the interaction term is of the additive type.

Using Eq.~\eqref{v+fjmn2p0}, Eq.~(\ref{E4}), and Eq.~(\ref{E5}), and recalling that $\rho_{\text{int}} = f(\rho_m, \varphi)$, we can express the components of the energy-momentum tensor as:
\begin{eqnarray}
    T^0_0 &=& -\rho_m - \widetilde{\zeta}(\rho_m) - \widetilde{C}_2, \\
    T^1_1 &=& T^2_2 = p_{\text{int}} - V(\varphi).
\end{eqnarray}
For simplicity, let's solve these using the barometric equation between the interaction pressure and interaction energy density:
\begin{eqnarray}
    p_{\text{int}} = \widetilde{w}_{\text{int}} \rho_{\text{int}},
\end{eqnarray}
and make a choice for $\widetilde{\zeta}(\rho_m)$:
\begin{eqnarray}
    \widetilde{\zeta}(\rho_m) = \rho_m.
\end{eqnarray}
One must note that this is the simplest choice for $\widetilde{\zeta}(\rho_m)$ which is dimensionally correct. One may put a numerical factor on the right hand side of the above equation, but in our case we assume that numerical factor is of the order of one.

Using these, we can solve the two Einstein equations which yields:
\begin{eqnarray}
    \rho_m (r) &=& \frac{1}{2} \left( \frac{1 - \lambda^2}{2 - \lambda^2} \right) \frac{1}{r^2} - \frac{1}{2} \widetilde{C}_2, \\
    \rho_{\text{int}} (r) &=& \frac{1}{\widetilde{w}_{\text{int}}} [T^1_1 + V(\varphi_0)], \label{jmn2_rho_p_soln_p0}
\end{eqnarray}
where $V(\varphi_0)$ is the constant value of the potential evaluated at $\varphi = \varphi_0$.
Using the first of the equations \eqref{jmn2_rho_p_soln_p0}, we can also obtain a solution for $\rho_{\text{int}}$ from equation \eqref{v+fjmn2p0}, which is:
\begin{eqnarray}
    \rho_{\text{int}} = \frac{1}{2} \left( \frac{1 - \lambda^2}{2 - \lambda^2} \right) \frac{1}{r^2} + \frac{1}{2} \widetilde{C}_2  - V(\varphi_0).
\end{eqnarray}
Equating the two expressions for $\rho_{\text{int}}$, we have:
\begin{eqnarray}
    T^1_1 - \frac{1}{2} \widetilde{w}_{\text{int}} \left( \frac{1 - \lambda^2}{2 - \lambda^2} \right) \frac{1}{r^2} =  \frac{1}{2} \widetilde{C}_2  - (1+\widetilde{w}_{\text{int}}) V(\varphi_0),\nonumber
\end{eqnarray}
from which we see that if we assume $\widetilde{w}_{\text{int}}$ is a constant, then the right hand side becomes a constant, while the left hand side is a function of $r$. To get a mutually consistent solution, we have to allow an $r$-dependent $\widetilde{w}_{\text{int}}$:
\begin{eqnarray}
    \widetilde{w}_{\text{int}} (r) = \frac{2[T^1_1 + V(\varphi_0)]}{\widetilde{C}_2 - 2 V(\varphi_0) - T^0_0}.
\end{eqnarray}
Henceforth we will assume that the fluid has exactly zero pressure. This is what we expected and we will try to see that whether such an assumption can produce a consistent result in the present case. We summarize the results:
\begin{eqnarray}
    p_m &=& 0, \\
    \varphi &=& \text{constant} = \varphi_0, \\
    \rho_m &=& \frac{1}{2} \left( \frac{1 - \lambda^2}{2 - \lambda^2} \right) \frac{1}{r^2} - \frac{1}{2} \widetilde{C}_2, \\
    \rho_{\text{int}} &=&  \frac{1}{2} \left( \frac{1 - \lambda^2}{2 - \lambda^2} \right) \frac{1}{r^2} + \frac{1}{2} \widetilde{C}_2  - V(\varphi_0),  \\
    p_{\text{int}} &=& \frac{1}{(2-\lambda^2)}\frac{1}{r^2}\left[\frac{(1-\lambda)^2 A - (1+\lambda)^2 B r^{2\lambda}}{A - B r^{2\lambda}}\right] + V(\varphi_0)\,.\nonumber\\ 
\end{eqnarray}
Using the above method, we can derive the solution for a non-minimally coupled scalar field for a spherically symmetric static spacetime in the present case. In these scenarios, since the scalar field becomes a constant, we can write the following expression of $\rho_\varphi$ and $p_\varphi$:
\begin{eqnarray}
       \rho_{\varphi} &=&  V(\varphi_0),\\
    p^{(r)}_{\varphi} &=& - V(\varphi_0),\\ p^{(\vartheta)}_{\varphi} &=&  p^{(\phi)}_{\varphi} = - V(\varphi_0).
\end{eqnarray}
Therefore, the scalar field component of the composite fluid behaves like the cosmological constant (with $\omega_{\varphi} = -1$). At the same time, from the above set of solutions we can see that $\omega_{m} = 0$, and consequently, the non-minimally coupled matter represents dark matter, with the interaction between them varying with radial distance. 
This results show that indeed we can get a set of consistent results if we assume that the matter part of the composite fluid has zero pressure. 
Of course, our assumption was based on phenomenological requirements and moreover, this assumption made the calculations straightforward. We summarize the nature of $\omega_\varphi$ and $\omega_m$ in Table \ref{table2} for both the minimal and non-minimal scenarios.

To consider the energy conditions in JMN-2 spacetime, we note from Eqs.~\eqref{f0soln22} and ~\eqref{f1soln22} that
\begin{align}
    \rho &= \left(\frac{1-\lambda^2}{2-\lambda^2}\right)\frac{1}{r^2}, \label{tot_rho_p_jmn-2} \\
    p &= \frac{1}{(2-\lambda^2)}\frac{1}{r^2}\left[\frac{(1-\lambda)^2 A - (1+\lambda)^2 B r^{2\lambda}}{A - B r^{2\lambda}}\right], \nonumber
\end{align}
where $A = \frac{(1+\lambda)^2 R_b^{\lambda-1}}{4\lambda \sqrt{2-\lambda^2}}$ and $B = \frac{(1-\lambda)^2 R_b^{-\lambda-1}}{4\lambda \sqrt{2-\lambda^2}}$.
We see from the above expressions, that the total energy density and  pressure in JMN-2 spacetime correspond to an ideal fluid which has isotropic pressure.  The parameter $\lambda$, which specifies the spacetime has the range $0 \leq \lambda < 1$, which immediately implies $\rho > 0$. It can be verified that the pressure decreases from the center towards the boundary at $R_b$ and becomes zero staying positive along the way, i.e., $p \geq 0$. So we have $\rho + p > 0$. Furthermore, we can calculate from Eq.~\eqref{tot_rho_p_jmn-2}, that the condition $\rho \geq \left|p \right| $ implies
\begin{align}
    \left(\frac{r}{R_b}\right)^{2\lambda} \geq - \left(\frac{1 + \lambda}{1 - \lambda} \right), \nonumber
\end{align}
which is always trivially satisfied, as the quantity in the right hand side of the inequality is  always negative. The positivity of the energy density and pressure also implies $\rho + 3p \geq 0$. The equality holds when $\lambda = 1$. So all the energy conditions, viz., weak, strong, null and dominant; are satisfied in JMN-2 spacetime. 

\begin{widetext}
\begin{table}
\captionsetup{width=\textwidth}
\renewcommand{\arraystretch}{1.7} 
\setlength{\tabcolsep}{3pt}      
\begin{tabular}
{|p{3.5cm}|p{2cm}|p{2cm}|p{2cm}|p{6.5cm}|}
\hline
Nature of coupling & $\omega_\varphi$ at $r\to 0$ & $\omega_\varphi$ at $r\to R_b$ & $\omega_m$ at $r\to 0$ & $\omega_m$ at $r\to R_b$ \\
\hline
Minimal coupling  
& $-1$ 
&$-1$ 
& $\frac{1-\lambda}{1+\lambda}$ 
& $\frac{1}{4\lambda}\left[(1+\lambda)^2 \Delta_1(\lambda, R_b) - (1-\lambda)^2 \Delta_2(\lambda, R_b) \right]$ \\
\hline
Non-minimal coupling 
& $-1$ 
& $-1$ 
& $0$ 
&$0$ \\
\hline
\end{tabular}
\caption{Nature of the equation of state parameters, $\omega_\varphi$ and $\omega_m$, of scalar field and matter, respectively, across different parameter spaces in the JMN-2 spacetime.}
\label{table2}
\end{table}
\end{widetext}
\section{Conclusion}
\label{four}

In this article, we discussed the possible end states of a gravitational collapse in a two-component system. No precise dynamical theory is there that gives a proper general relativistic (GR) description of the process in which one obtains a spherically symmetric, static spacetime as the end product of a gravitational collapse. It is assumed that the end state of gravitational collapse, of the top-hat type, ends in a virialized state. Analytically it is still not understood how this virialization happens if one uses GR techniques. Consequently, we have not tried to formulate a detailed theory following which a collapsing system with two components of matter forms a final spherically symmetric, static space-time.
On the other hand, if a collapsing system has to yield galaxies and clusters of galaxies as the end state then we must have some form of a stable spacetime at the end of the collapse. In this paper, we have assumed that this end phase is a static, spherically symmetric spacetime. This assumption makes the calculations analytically doable and infers some light on the possible end states of gravitational collapse.

In general, it is difficult to have a pure scalar field seeding a static, spherically symmetric spacetime. There are of course some solutions and perhaps the most popular of those solutions is the JNW (Janis-Newman-Winicour) solution, which involves a massless scalar field \cite{Janis:1968zz, Vir1}. It is also known that one cannot have sourceless scalar field solutions outside the black hole horizon. It turns out that the scalar field accompanied by a relativistic ideal fluid can seed static, spherically symmetric spacetimes. In this case, the fluid helps the scalar field to stabilize. This fact compelled us to investigate the end stages of gravitational collapse in the presence of two components. One of these components is assumed to be a scalar field and it is assumed to be the dark energy constituent. The other component is supposed to be a fluid with nearly vanishing pressure and this component signifies the dark matter part. Our main point of investigation is related to the finding out the asymptotic, stable end phase of a gravitational collapse where the spacetime is seeded by the remnants of the dark sector components. It is known that most of the stable gravitationally bound states in the astrophysical scales do contain dark matter. Beyond a certain cosmological scale, dark energy can significantly influence the structure formation of dark matter. In the introduction (Sec.~\ref{sec0}), we discussed existing models for the formation of dark matter structures in the presence of dark energy. In this paper, we demonstrate that the properties of the composite fluid, consisting of both dark matter and dark energy, can mimic those of composite fluids in the clustered dark energy models. However, because we examine the properties of this two-component system in the context of a static, spherically symmetric spacetime—representing the final equilibrium state of gravitational collapse, under specific initial conditions, this resemblance holds only in the final static equilibrium state.

In the present work, we have assumed that the components of the remnants of the dark sector (fluid representing dark matter and a scalar field representing dark energy candidate) may act as two non-interacting components or may interact non-minimally with each other. In the latter case, the dark sector components directly exchange energy between themselves. It is found that the nature of the EoS of the two components of the dark sector in the stable asymptotic end state may differ from their corresponding values in the cosmological scale. This is the primary reason why we call the two components seeding the end stable spacetime as remnants of the original dark sector components. In the present case, it is important to note that during the collapsing phase, we consider a contracting general collapsing spacetime (Eq.~\ref{sptcollapse1}) with specific regular initial data, and in the asymptotic end state, this evolves into a spherically symmetric, static spacetime. Our analysis focuses on the properties of the dark matter and dark energy-like scalar fields at the final equilibrium state; we did not explore the dynamical evolution of their characteristics. The mutual interaction between the dark sectors during the gravitational contraction phase may cause dark matter and dark energy to behave somewhat differently than they do in the background, and this difference persists up to the asymptotic equilibrium state.

To concretely study the properties of the matter and scalar field composition at the asymptotic end equilibrium state of a gravitational collapse, we assume that the end-state spacetimes are represented by either the JMN-1 or JMN-2 spacetimes. Both of these spacetimes can be shown to represent the end equilibrium states of gravitational collapse \cite{Joshi1, Joshi2}. Both these spacetimes have a central naked singularity and both of them can be smoothly joined to an external Schwarzschild spacetime. Both of these spacetimes are specified by some set of real parameters whose values can be inferred from phenomenological requirements. These parameters in general appear in the equations of states of the components present in the spacetimes. It is seen that in almost all of the cases we discuss the scalar field sector invariably has a negative equation of state whose highest value may be $-1.34$ in the case of JMN-1 spacetime. It is seen that in presence of dark matter remnants one cannot have a canonical scalar field in the end state, where the end state is the JMN-1 spacetime.
The scalar field in the JMN-1 spacetime (whether minimally coupled or non-minimally coupled) is always a phantom like scalar field. On the other hand it is seen that JMN-2 spacetime cannot distinguish between a canonical scalar field or a phantom like scalar field, both kind of fields can act as the dark energy remnant in the end state. For the case of JMN-2 spacetime one can have the scalar field equation of state exactly around $-1$. These results show that the scalar field in these spacetimes, when accompanied by a zero pressure or low-pressure fluid, naturally behaves as some remnant of the cosmological dark energy and in some cases, it exactly behaves as a cosmological constant. The matter component can also have various equations of state. In a spherically symmetric, static spacetimes, most of the time the equations of state are functions of the radial coordinate (unless when they are constants). We have seen that in JMN-1 spacetime it is very difficult to have the matter equation of state to be exactly near zero such that the scalar field equation of state remains near $-1$ simultaneously. This fact shows that the equilibrium process may induce changes in the equations of state of the components of the dark sector. In the case of JMN-2 spacetime, one may reduce the matter pressure but not exactly to zero for the case where the components are not interacting with each other. 

In all the cases discussed, we have examples of clustered dark energy where dark energy and dark matter are virialized together. The two dark sectors may have a nontrivial interaction and this interaction may produce other interesting phenomena as the phenomena related to the fifth force. In this article we do not discuss the fifth force effects, we intend to discuss this issue in a later publication.

\section{Acknowledgements}
Debanjan Debnath would like to express his sincere gratitude to the Council of
Scientific and Industrial Research (CSIR, India, CSIR Award No.: CSIRAWARD/SPM-JR2022/13184) for funding the work.


\end{document}